\documentclass[twocolumn,         % Format : preprint, twocolumn
               showpacs,longbibliography,            % Pacs : showpacs, noshowpacs
               showkeys,preprintnumbers,     % Preprint: preprintnumbers,
               aps,                 % Society: ...
               prd,          	    % Journal Style : pra, prb, prc, prd, pre,
               letterpaper,             % Size : a4paper, ...
               superscriptaddress,      % Affiliation (Title) : groupedaddress,
               nofootinbib,         % Footnote: footinbib, nofootinbib
               tightenlines,        % Remove additional spaces in a line
               floats,floatfix      % Floating pictures and tables
               ]{revtex4-1}
\usepackage{physics}
\usepackage{epsf}
\usepackage{graphicx,color}
\usepackage{subfigure}
\usepackage{latexsym}
\usepackage{amsmath,amssymb}        
\usepackage[colorlinks=true,linkcolor=blue,citecolor=blue]{hyperref}
\usepackage{mathrsfs}
\usepackage{comment}
\usepackage{soul}
\usepackage{cancel}
\definecolor{purple}{rgb}{0.58,0.0,0.83}
\definecolor{orange}{rgb}{1,0.5,0}
\DeclareSymbolFontAlphabet{\mathrsfs}{rsfs}
\DeclareMathAlphabet{\mathcal}{OMS}{cmsy}{m}{n}

\begin{document}
% -----> TITLE 

\title{Effects of Self-Interaction and of an Ideal Gas in Binary Mergers of Bosonic Dark Matter Cores}
% ----->   AUTHORS   <-----

\author{Carlos Tena-Contreras}
\email{1423327c@umich.mx}
\affiliation{Instituto de F\'{\i}sica y Matem\'{a}ticas, Universidad
              Michoacana de San Nicol\'as de Hidalgo. Edificio C-3, Cd.
              Universitaria, 58040 Morelia, Michoac\'{a}n,
              M\'{e}xico.}           

\author{Iv\'an  \'Alvarez-Rios}
\email{ivan.alvarez@umich.mx}
\affiliation{Instituto de F\'{\i}sica y Matem\'{a}ticas, Universidad
              Michoacana de San Nicol\'as de Hidalgo. Edificio C-3, Cd.
              Universitaria, 58040 Morelia, Michoac\'{a}n,
              M\'{e}xico.}               

\author{Francisco S. Guzm\'an}
\email{francisco.s.guzman@umich.mx}
\affiliation{Instituto de F\'{\i}sica y Matem\'{a}ticas, Universidad
              Michoacana de San Nicol\'as de Hidalgo. Edificio C-3, Cd.
              Universitaria, 58040 Morelia, Michoac\'{a}n,
              M\'{e}xico.}  

\author{Jens Niemeyer}
\email{jens.niemeyer@phys.uni-goettingen.de}
\affiliation{Institut f\"ur Astrophysik und Geophysik, Georg-August-Universit\"at G\"ottingen, D-37077 G\"ottingen, Germany}
% --->   DATE

\date{\today}

% -----> ABSTRACT

\begin{abstract}
We study binary mergers of dark matter cores in the Bose-Einstein condensate (BECDM) model. We include two scenarios: scalar self-interaction and the presence of a gravitationally coupled ideal gas. Using 3D simulations of the Gross-Pitaevskii-Poisson and Schr\"odinger-Poisson-Euler systems, we analyze the properties of the resulting remnants. We find that the final core-mass ratio reaches a stable average value after the merger. Repulsive self-interaction increases the mass of the final solitonic core, while attractive interaction enhances mass loss. In mergers involving an ideal gas, namely of fermion-boson stars, a stable solitonic core always forms in the bosonic component, even when the gas dominates, whereas the gas itself does not form a compact core. We explain these results using energy scalings and find that without self-interaction, equilibrium cores follow $E \propto -M^3$, which leads to an almost universal merger fraction. Self-interaction changes this scaling, because repulsive $g$ moves the system toward a milder $E \propto -M^2$ scaling and increases mass retention, while attractive $g$ strengthens binding and favors mass ejection. In the case of interaction with an ideal gas, this component only modifies the gravitational background and does not change the intrinsic scaling of the bosonic part. These results show that the merger outcome is not universal but controlled by the interaction strength, while solitonic BECDM cores remain robust across diverse environments including gas.
\end{abstract}

% ----->   PACS

%\pacs{keywords: self-gravitating systems -- dark matter -- Bose condensates}
%07.05.Tp Computer modeling and simulation
%07.05.Mh Neural networks, fuzzy logic, artificial intelligence
%05.45.Tp Time series analysis
%04.30.-w Gravitational waves

% ----->   MAKETITLE   <-----

\maketitle

%----------------------------------------------------------%
% ---------------->   INTRODUCTION     --------------------% 
% ------------------------------- -------------------------%
\section{Introduction}
\label{sec:intro}

Ultralight or fuzzy dark matter (FDM) models describe the dark sector as a coherent scalar field with an extremely small particle mass, typically \( m \sim 10^{-23}\text{--}10^{-21}\,\mathrm{eV} \)
\cite{Matos:2000ss,Hu:2000ke,Hui:2016}. Structure formation simulations of FDM based on the solution of the Schr\"odinger--Poisson (SP) system that governs the dynamics of this dark matter field, indicate that at galactic scales the evolution of structures collapses to form solitonic cores surrounded by extended 
NFW-like halos (e.g. \cite{Schive:2014dra,Mocz:2017wlg,Veltmaat_2018,Desjacques:2017fmf,Gotinga2022}). These cores correspond to the ground-state equilibrium of the SP equations \cite{GuzmanUrena2004}, and play a central role in the small-scale phenomenology of FDM halos, providing cored density profiles surrounded by an envelope with characteristic interference patterns.

The robustness of these solitonic configurations under gravitational interactions
has been extensively studied, from very basic tests that include the resistance to perturbations, its attractor properties in non-spherical scenarios \cite{BernalGuzman2006b}, genuine solitonic behavior under collisions \cite{BernalGuzman2006a}, until the aforementioned Structure Formation Simulations that show their universality and inherence to the FDM model.

In between the basic and state of the art simulations there is an interesting scenario that calls the attention, the binary merger of solitons, namely the ground-state solutions of the SP system. In \cite{PAREDES201650} a general scenario of mergers, along with a wide exploration of the parameter space was presented, including the effects of phase and degrees of freedom of the system on the behavior of the resulting fused structure.

Going further, in \citet{Schwabe:2016} a controlled set of 3D simulations of binary core mergers leads to the analysis on the relaxation process of the merger, with an intriguing result, namely, that the mass of the final virialized core $M_{c,\mathrm{final}}$ after gravitational relaxation, is a nearly universal fraction of the addition of the core mass of the initial two solitons prior to mergeing $M_{c,\mathrm{initial}}=M_{c,1}+M_{c,2}$, with \( M_{c,\mathrm{final}} \simeq m\,M_{c,\mathrm{initial}} \), and $m$ is a magic factor between 0.6 and 0.7 depending on the use of isolation or periodic boundary conditions. 
This result was shown to hold over  a wide range of mass ratios and initial angular momentum. The mass deficit is attributed to gravitational cooling \cite{SeidelSuenCooling,GuzmanUrena2006}, consisting of the emission of the excess of kinetic energy outside the final core as a process of relaxation and aiming the condensation of the core. For this, absorption of radiated modes by the boundaries in simulations is fundamental, in order to allow the final core to relax free of interference coming from the outside as illustrated to happen when using non-absorbing boundary conditions, and happens similarly when using periodic boundary conditions although part of the halo is never removed from the numerical domain (see \cite{periodicas} for details and examples). Thus, a fraction of the bosonic mass is ejected into a diffuse halo during the violent relaxation phase.

Also important are the effects of self-interaction, which lead to differences in the merger process and final configuration in terms of robustness and radiative response of the final configuration \cite{PhysRevD.104.083532,TanjaBinary}.
In this work we extend the results found in \citet{Schwabe:2016}, by exploring two generalizations of physical relevance:
(i) the inclusion of a \emph{self-interaction} term, so that the dynamics is governed by the Gross--Pitaevskii--Poisson (GPP) system, which is the SP system with self-interaction, and (ii) the coupling of the scalar field to an \emph{ideal gas} component.
Repulsive quartic self-interactions (\(\propto g|\psi|^4\)) have been shown to increase core sizes and decrease central densities, which adds to the quantum-pressure even in the 
Thomas--Fermi regime (e.g. \cite{TanjaBinary,Suarez:2016eez,https://doi.org/10.48550/arxiv.2301.10266}). Because the internal structure and oscillation spectra of the cores depend on \(g\), we expect the efficiency of gravitational cooling \cite{SeidelSuenCooling,GuzmanUrena2006}, and the final-to-initial core mass ratio to be modified in mergers of cores with self-interaction.

On the other hand, coupling the bosons to an ideal gas adds an additional energy-exchange channel. During a merger, part of the bosons' kinetic energy may be transferred to the gas component through drag or dissipation, altering the amount of mass expelled into the halo. Conversely, shock heating and pressure gradients in the gas may perturb the merged core and enhance mass loss. These effects are unexplored in controlled GPP$+$gas simulations, and are astrophysically motivated by environments where baryonic or thermal components coexist with the dark matter core \cite{FermionBosoStars2024}.

We therefore address the analysis of how the final-to-initial core mass ratio \( M_{c,\mathrm{final}} / M_{c,\mathrm{init}} \) depends on the self-interaction strength \( g \) and on the presence of an ideal gas. 
For this, we perform 3D simulations of binary mergers of stationary solutions, in the first case of solitonic solutions with self-interaction. In the second case, we merge Newtonian Fermion-Boson Stars (FBS) \cite{Alvarez_Rios_2023}, that is, stationary solutions of the Schr\"odinger-Newton-Euler system of equations, which are the Newtonian version of the stationary solutions of Einstein's equations sourced by a complex scalar field minimally coupled to a perfect fluid \cite{HENRIQUES198999,HENRIQUES1990511}.

We find that the nearly universal final-to-initial core mass ratio reported in \cite{Schwabe:2016} reflects the scaling properties of equilibrium configurations. In the absence of self-interaction, the solitonic cores obey a mass-radius relation $r_c \propto 1/M_c$, implying an equilibrium energy scaling $E \propto -M_c^3$, which leads to a universal merger fraction. When a quartic self-interaction is included, the additional interaction energy modifies the effective mass scaling of equilibrium configurations, shifting the retained core fraction so that  repulsive interactions drive the system toward a regime closer to $E \propto -M_c^2$ and increase the mass retention, whereas attractive interactions trigger mass loss. In contrast, when an ideal gas component is present, the coupling is only gravitational and the gas remains more spatially extended than the bosonic core, and its effects reduce to reshaping the background potential without altering the intrinsic cubic scaling of the bosonic component. These energetic scaling arguments provide a unified interpretation of the numerical trends observed in the simulations.

Our results extend those in \cite{Schwabe:2016} related to the merger relation of initial vs final core mass, into the regimes of self-interaction and coupling to the effects of a gravitationally coupled ideal gas. 
They offer insight into how core condensation and mass loss proceed in more complex environments, and contribute to refined models of core--halo structure and evolution in bosonic dark matter structures.

The paper is organized as follows. In Section \ref{sec:self-interaction} we present the case of mergers of solitons with self-interaction, in Section \ref{sec:ideal-gas} the merger of FBS and in Section \ref{sec:conclusions} we draw some conclusions.

%----------------------------------------------------------%
% ---------------->   Section     --------------------% 
% ------------------------------- -------------------------%
\section{Effects of self-interaction}
\label{sec:self-interaction}

% -----------
\subsection{Equations}

In this scenario, the system that governs the dynamics of the bosonic cloud  is the GPP with self-interaction:

\begin{eqnarray}
    i\partial_{t} \Psi = \left(-\dfrac{1}{2}\nabla^2 + V + g|\Psi|^2\right)\Psi, \label{eq:Schrodinger} \\
    \nabla^2 V = |\Psi|^2, \label{eq:Poisson}
\end{eqnarray}

\noindent where $\Psi$ is the complex order parameter that represents the macroscopic wave function describing the BEC dark matter dynamics. The quantity $|\Psi|^2$ corresponds to the bosonic cloud density, while $g$ is the self-interaction constant.

% -----------
\subsection{Initial Conditions for the merger of cores}

Cores are ground state solutions of the GPP system, which are the nodeless stationary solutions of the above system, commonly referred to as Newtonian Boson Stars \cite{Ruffini:1969,GuzmanUrena2004,GuzmanUrena2006,CarlosIvanFranciscoGA}, which are very robust and stable \cite{GuzmanUrena2006,BernalGuzman2006b,ChengNiemeyer2021,AlvarezGuzmanMadelung,CarlosIvanFranciscoGA}. 

\begin{figure}
    \centering
    \includegraphics[width=\linewidth8]{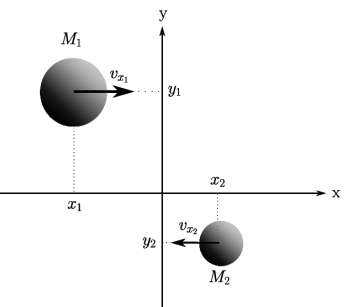}
    \caption{Illustration of the initial setup for the binary merger of soliton cores.}
    \label{fig:binary_merger_illustration}
\end{figure}

\textit{Preparation of the Binary System.} Each BS is characterized by the central value of its wave function, $\psi_c$, which determines its mass and size. An illustration of the initial conditions appears in Fig. \ref{fig:binary_merger_illustration}. In order to introduce asymmetry between the two BSs, we define a dimensionless scaling factor $\lambda$, that controls the relative amplitude, position, and velocity of the soliton at the right with respect to the soliton at the left. The initial central value of the wave function of the left configuration is \(\psi_{c,1}\), while that at the right configuration is \(\psi_{c,2} = \lambda^2 \psi_{c,1}\).  Each soliton is initially centered at $\vec{x}_1 = (x_1,y_1,0)$ and $\vec{x}_2 = -\lambda \vec{x}_1$, with the center of mass at the coordinate origin.
In this way, the total bosonic wave function is constructed by superposing both wave packets, each boosted with its respective velocity:

\begin{equation}
    \Psi(0, \vec{x}) = \psi_1 e^{i\vec{v}_1 \cdot \vec{x}} + \psi_2 e^{i\vec{v}_2 \cdot \vec{x}}.
    \label{eq:initial_order_parameter_BS}
\end{equation}

\noindent where \( \vec{v}_1 = (v_{x,1}, 0, 0)\) and \(\vec{v}_2 = -\lambda \vec{v}_1\),
while $\psi_1$ and $\psi_2$ represent the order parameter of the soliton at the left and at the right, respectively.

% -----------
\subsection{Numerical Scheme and Simulations Setup}

The simulations are performed using the CAFE-FDM  code \cite{periodicas}, 
where the GPP system is solved using a third-order Runge-Kutta method for 
time integration, while the Gross-Pitaevskii equation (\ref{eq:Schrodinger}) and 
Poisson equation (\ref{eq:Poisson}) are discretized using the Fast Fourier 
Transform.

Boundary conditions for $\Psi$ are periodic, while for the gravitational potential, Dirichlet boundary conditions are imposed by setting \( V = 0 \) at the boundaries of the domain. This assumption isolates the system gravitationally from external sources. As a result, the binary collisions resemble those of an isolated system, even though the matter field evolves under periodic conditions.

The simulations are carried out in a cubic spatial domain \( D = [-20, 20]^3 \), discretized using \( N = 128 \) grid points in each direction, corresponding to a spatial resolution of \( h = 0.3125 \). The system is evolved over 250 time units with a  resolution of \( \Delta t = 0.01 \), which satisfies the Courant condition \( \Delta t / h^2 < \dfrac{1}{6 \pi^2} \), as recommended in \cite{ChengNiemeyer2021}.

\subsection{Diagnostics}

In order to analyze the evolution of the system during and after the merger, we monitor the ratio between the masses of the initial cores and the resulting final core. This allows us to reproduce and eventually generalize the findings of \cite{Schwabe:2016}, related to the magic fraction between the final solitonic mass and the sum of the initial soliton masses.  

For this we need the angle-averaged density profile of a function $f$ with respect to the coordinate origin after the merger, defined as:

\begin{equation}
    f_{\text{avg}}(t, r) = \dfrac{1}{4\pi} \int_\Omega f \, d\Omega,
    \label{eq:average_density}
\end{equation}

\noindent where $f=|\Psi|^2$ for the merger of BEC solitons. The solid angle is given by $\Omega = [0, \pi] \times [0, 2\pi]$, with the differential element $d\Omega = \sin\theta \, d\theta \, d\phi$. Once a final core has formed as a result of the  merger, the final core mass is computed as:

\begin{equation}
M_{c,{\rm final}}(t) = 4\pi \int_0^{r_c} f_{\text{avg}}(t,r) r^2 \, dr,
\label{eq:core_mass_merger}
\end{equation}

\noindent where $r_c$ is the core radius. The merged core mass $M_{c,{\rm final}}$ is then compared to the initial masses of the individual initial cores, $M_{c,1}$ and $M_{c,2}$, through the core mass ratio:

\begin{equation}
    R_{M_c} = \dfrac{M_{c,{\rm final}}}{M_{c,1} + M_{c,2}},
    \label{eq:mass_core_ratio}
\end{equation}

\noindent which is the final versus initial core mass ratio.

Now, the density profile of a soliton, namely the ground 
state solution of the GPP system \cite{GuzmanUrena2006} is approximated
with the empirical formula 

\begin{equation}
\rho_c(r)=\rho_0\left[
1 + 0.091 \left(\frac{r}{r_c}\right)^2
\right]^{-8},
\label{eq:soliton_profile}
\end{equation}

\noindent found in \cite{Schive:2014dra}, within structure formation simulations.
In this formula $\rho_0$ is the central density of the core whereas 
$r_c$ is the core radius. The core radius is considered to be
the radius at which the density is half the peak density,
and is the concept used in \cite{Schwabe:2016}, where 
the magic factor $m\sim 0.7$  between final and initial core masses was 
found. 

Unlike in that original analysis, where isolation boundary
conditions were used via the implementation of a sponge, in our 
analysis we use periodic boundary conditions which represent a different
physical system because the excess of kinetic energy expected to be
ejected through the gravitational cooling \cite{SeidelSuenCooling,GuzmanUrena2006} reenters the numerical 
domain and consequently the cores remain surrounded within a hyperkinetic
sea of bosons. This changes slightly the magic number from $m\sim 0.7$ to $m\sim 0.6$ for the
base case of fuzzy dark matter ($g=0$), as seen
below, and we confirm that the cored mass ratio
certainly gives a nearly constant value across a number of initial 
conditions.

\subsection{Results of simulations}

To study the effect of the self-interaction on the compactness and mass distribution of soliton cores, formed during mergers with non-zero angular momentum, we vary the self-interaction constant with values $g = -0.1$, $0.0$, $0.1$, $0.2$ and $0.3$. The initial conditions for the soliton at
the left are fixed to $\psi_{c,1} = 1$, with initial position $\vec{x}_1 = (-5, y_1, 0)$, where the impact parameter takes values $y_1 = 5$ and $10$. The non-zero velocity component takes values 
$v_{x_1} = 0.05$, $0.1$, $0.15$, $0.2$, $0.25$. For the configuration at the right we use the 
scale factor values $\lambda =  \sqrt{1.5}$ and $\sqrt{2.0}$, in order to cover at least two mass ratios between the two initial configurations. The choice of scaled position and velocity ensures that the center of mass of the system remains near the origin of the domain during the evolution.

\begin{figure}
    \centering
    \includegraphics[width=\linewidth]{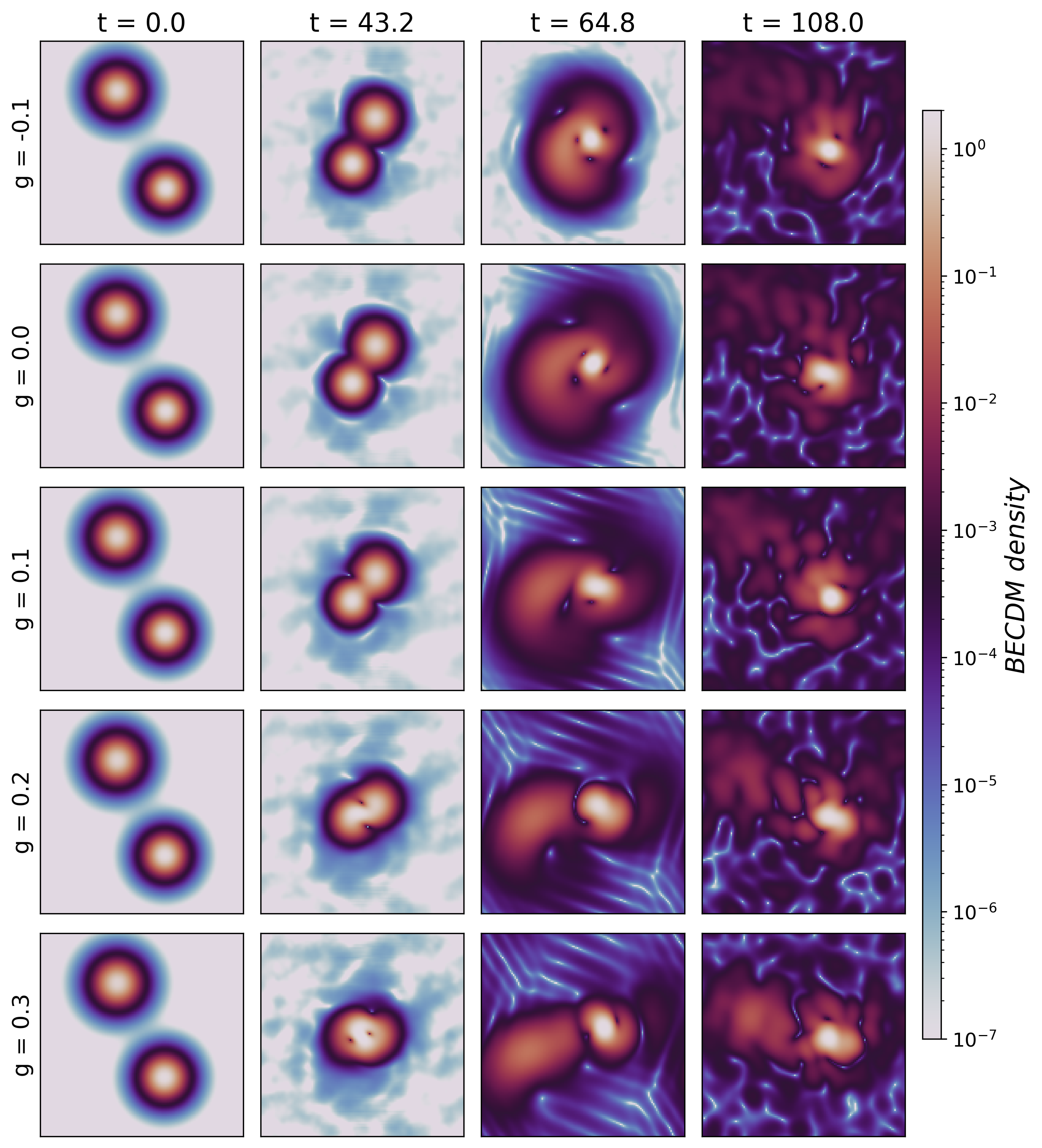}
    \caption{Snapshots of the density evolution on the plane \(z = 0\) at various times. Each row corresponds to a different value of the self-interaction  \(g = -0.1\), 0, 0.1, 0.2, 0.3. The parameters for these simulations are \(\lambda = \sqrt{2.0}\), \(y_1 = 10\), and \(v_{x,1} = 0.2\).}
    \label{fig:evolution_self_interaction}
\end{figure}

 \textit{Core Formation.} Figure~\ref{fig:evolution_self_interaction} shows the evolution of the density on the plane \( z = 0 \) for the specific case with initial conditions \(\lambda = \sqrt{2.0}\), \(y_1 = 10\) and \(v_{x_1} = 0.2\). Snapshots are taken at times \(t = 0\), 43.2, 64.8, and 108.0 for various values of the self-interaction constant \(g = -0.1\), 0, 0.1, 0.2, and 0.3. 

A stable core becomes distinguishable at around \( t \sim 64 \). As the system continues to evolve, a diffuse halo forms around the central core, which resembles gravitational cooling (see \cite{SeidelSuenCooling} for \(g = 0\) and \cite{GuzmanUrena2006} for \(g \neq 0\)) of the core. The resulting configuration exhibits the dense central solitonic core embedded within a slowly decaying halo. These solitonic cores remain stable for all values of $g$, confirming consistency with
earlier studies \cite{GuzmanAlvarezGonzalez2021, Schwabe:2016, periodicas}.

We then apply the empirical density profile to fit the soliton with self-interaction proposed 
in \cite{CarlosIvanFranciscoGA}, that generalizes (\ref{eq:soliton_profile}), 
and is slightly different from  that in \cite{ChengNiemeyer2021}:

\begin{equation}
    \psi^2_{\text{soliton}}(r) = \psi_c^2 \left[1 - \left(2^{1/8} - 1\right) \left(\dfrac{r}{r_c}\right)^{2 + \beta} \right]^{-8},
    \label{eq:soliton_density}
\end{equation}

\noindent with
\begin{eqnarray}
    r_c = 1.306\psi_c^{-1/2} \left( 1 + a_1 \alpha + a_2 \alpha^2 \right), \\
    \beta = b_1\alpha + b_2\alpha^2 + b_3 \alpha^{1/2},
\end{eqnarray}
\noindent where \(\alpha = g\psi\), and 
\begin{eqnarray}
a_1 &=&  0.3681  \pm 24.50 \times 10^{-5}, \nonumber \\
a_2 &=&  0.0905  \pm 31.34 \times 10^{-5}, \nonumber \\
b_1 &=&  0.2842  \pm 10.71 \times 10^{-5}, \nonumber \\
b_2 &=&  0.0845  \pm 21.80 \times 10^{-5}, \nonumber \\
b_3 &=& -0.0117  \pm 5.443 \times 10^{-5}. \nonumber 
\end{eqnarray}

\noindent The profile depends on the central wave function \(\psi_c\) and the core radius \(r_c\), both modulated by the self-interaction. 

\begin{figure}
    \centering
    \includegraphics[width=7.5cm]{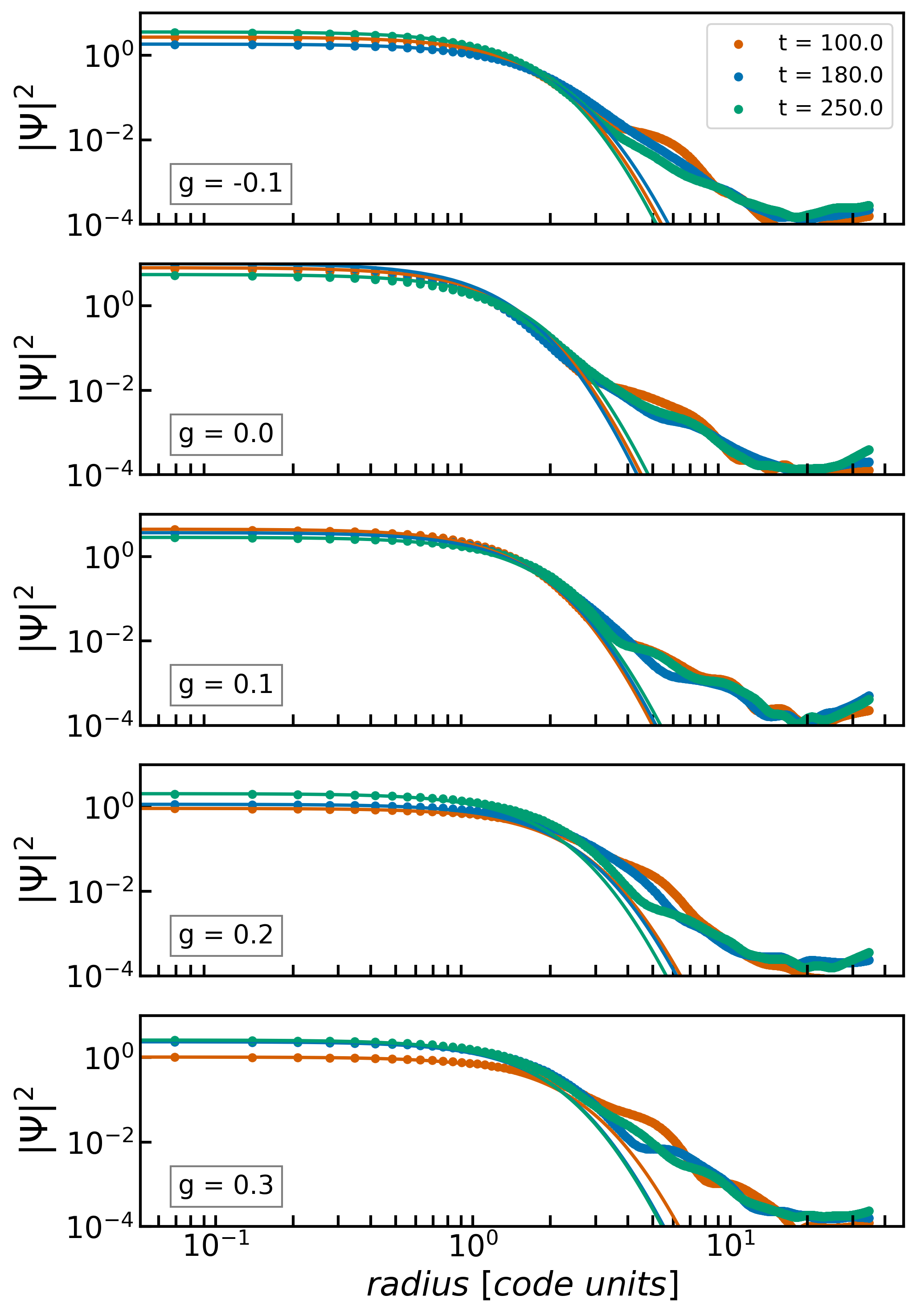}
    \caption{Angle-averaged density profile (\ref{eq:average_density}) 
shown with points, at three different times along with the empirical fit (\ref{eq:soliton_density}) with solid line. Each panel corresponds to a different value of \(g = -0.1\), 0, 0.1, 0.2, and 0.3. Here we use the parameters \(\lambda = \sqrt{2.0}\), \(y_1 = 10\), and \(v_{x_1} = 0.2\).}
    \label{fig:avg_and_fit}
\end{figure}

Figure~\ref{fig:avg_and_fit} shows the angle-averaged density profiles at three different times \(t = 100\), 180 and 200, along with fits that use Eq.~\eqref{eq:soliton_density}, for the case \(\lambda = \sqrt{2.0}\). The agreement between simulation and the empirical model in the core region confirms the emergence and stability of the solitonic core for each considered value of \(g\). 

 \textit{Core Mass Ratio.} In order to investigate the role of self-interaction on the final core mass with respect to the initial one, we compute the core mass ratio using Eq.~\eqref{eq:mass_core_ratio}, focusing on the post-merger regime once the final soliton has formed. 

\begin{figure}
    \centering
    \includegraphics[width=7.5cm]{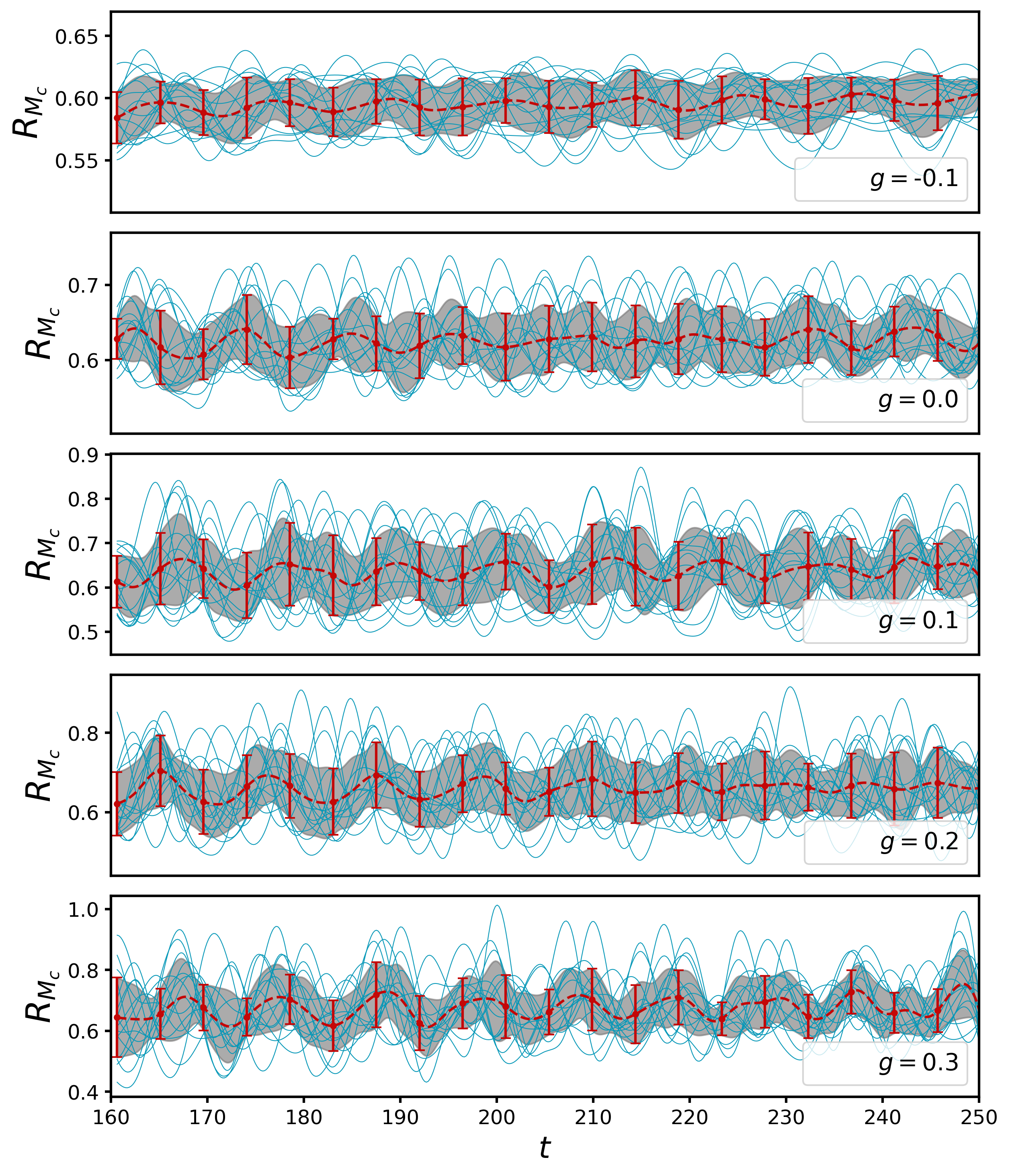}
    \caption{Time evolution of the core mass ratio \(R_{M_c}\) for 
\(g = -0.1\), 0.0, 0.1, 0.2, and 0.3, from top to bottom. Each panel 
shows results from 20 simulations for each $g$. The thick dotted line indicates the average value of \(R_{M_c}\) across simulations 
and the gray band its standard deviation.} 
    \label{fig:mass_core_ratio}
\end{figure}

Figure~\ref{fig:mass_core_ratio} contains the evolution of the ratio \(R_{M_c}\) in Eq. (\ref{eq:mass_core_ratio}) for different values of \(g\), and the 20 different simulations per case, namely two values of $y_1$, five values of $v_{x_{1}}$ and two values of $\lambda$. The plots indicate that \(R_{M_c}\) fluctuates around an average \(R_{M_c,\text{avg}}\) for each value of $g$, which are presented in Table~\ref{tab:mass_core_ratio}. As the self-interaction strength increases, the average mass retained in the final soliton core also increases, indicating that the repulsive self-interaction reduces the mass loss from the core during cooling/relaxation in consistency with the results in \cite{TanjaBinary}.

\begin{table}
   \centering
   \begin{tabular}{|c|c|c|c|c|c|}
   \hline
   \(g\) & -0.1 & 0.0 & 0.1 & 0.2 & 0.3 \\
   \hline
%   \(R_{M_c,\text{max}}\) & 0.6263 & 0.6866 & 0.7275 & 0.7697 & 0.8263 \\
%   \hline
   $\langle R_{M_c,\text{avg}} \rangle $ & 0.5636 & 0.6011 & 0.6217 & 0.6579 & 0.6839 \\
   \hline
   $\langle \sigma \rangle$ & 0.019 & 0.039 & 0.070 & 0.075 & 0.085 \\
   \hline
%   \(R_{M_c,\text{min}}\) & 0.4921 & 0.5055 & 0.5075 & 0.4387 & 0.4387 \\
%   \hline
   \end{tabular}
   \caption{
Average in time of $R_{M_c}$ of the plots in Fig.
\ref{fig:mass_core_ratio}, as well as
the average of the standard deviation $\sigma$ in time, calculated using the 20 simulations for 
each value
of $g$.}.
   \label{tab:mass_core_ratio}
\end{table}

Average in time of $R_{M_c}$ of the plots in Fig.
\ref{fig:mass_core_ratio}, as well as
the average of the standard deviation $\sigma$ in time, are shown in Table 
\ref{tab:mass_core_ratio}.
These results show the role of self-interaction in regulating the mass retention by the core. For \(g > 0\), repulsive interaction suppresses mass loss, stabilizes the final core, and enhances its compactness. In contrast, attractive interaction (\( g < 0\)) leads to greater mass dispersion into the halo, indicating a less efficient mass retention process. 
This is reflected in the growth of $R_{M_{c}}$ with $g$ in the Table.

\subsection{Explanation of the results}

So far we have obtained the final-to-initial core mass ratio based on numerical simulations for a given parameter space, and proceed now to draw an explanation of these results. For this we explore first the simplest case without self-interaction and later on  construct a generalization.

\subsubsection{Case $g=0$}

Let us consider the $g=0$ case and seek a physical explanation for the emergence of a nearly universal final core mass fraction in binary mergers. For a ground-state solution of the GPP system, which is virialized, the total energy satisfies
%\begin{equation}
$E = K + W = -K = \frac{1}{2}W$.
%\end{equation}
 Combining this with the scaling relation between core mass and radius
%\begin{equation}
$M_c\, r_c = \mathrm{const}$,
%\end{equation}
 and using $K \sim M_c/r_c^2$ together with $r_c \propto M_c^{-1}$, one finds
\begin{equation}
E_c \propto - M_c^3 .
\label{eq:Ecubic}
\end{equation}

\noindent
Consider now two initially far away solitons of masses $M_{c,1}$ and $M_{c,2}$. Neglecting their mutual gravitational interaction at large separation, the total initial energy scales as

\begin{equation}
E_i \sim -M_{c,1}^3 - M_{c,2}^3 .
\label{eq:Einitial}
\end{equation}

\noindent During the merger process, gravitational cooling ejects positive-energy bosonic modes to large radii, which implies that the final solitonic energy is bigger than the initial one

\begin{equation}
|E_{\rm final}| > |E_{\rm initial}| ,\nonumber
\end{equation}

\noindent since the system loses positive energy it becomes more bound. Now, assuming first that  the total final binding energy is contained in the relaxed solitonic core

\begin{equation}
E_{\mathrm{sol,final}} = E_{\mathrm{tot,final}} ,\nonumber
\end{equation}

\noindent and using the cubic scaling relation \eqref{eq:Ecubic}, we find that

\begin{equation}
M_{c,{\rm final}}^3 \gtrsim M_{c,1}^3 + M_{c,2}^3 .\nonumber
\end{equation}

\noindent Now, introducing the mass ratio $\lambda = M_{c,2}/M_{c,1}$, we obtain

\begin{equation}
\frac{M_{c,{\rm final}}}{M_{c,1}+M_{c,2}} \gtrsim \frac{(1+\lambda^3)^{1/3}}{1+\lambda},\nonumber
\label{eq:lambdarelation_new}
\end{equation}

\noindent which for equal-mass mergers ($\lambda = 1$) gives a key number
\begin{equation}
\frac{M_{c,{\rm final}}}{M_{c,1}+M_{c,2}}=2^{-2/3}\simeq 0.63 .
\label{eq:063}
\end{equation}

\noindent However, the assumption $E_{\mathrm{sol,final}} = E_{\mathrm{tot,final}}$ corresponds to the limiting case in which no bound non-solitonic component remains after relaxation. In realistic mergers, part of the density remains gravitationally bound although in a diffuse halo, therefore

\begin{equation}
|E_{\mathrm{sol,final}}| < |E_{\mathrm{tot,final}}| ,\nonumber
\end{equation}

\noindent and Eq. \eqref{eq:063} must be interpreted as an upper bound on the final core mass fraction.

This interpretation is consistent with the soliton--halo band discussed in \cite{Blum2025JCAP}, where the ``1/3 relation'' is defined 

\begin{equation}
M_{\mathrm{sol}} \propto |E_{\mathrm{tot}}|^{1/3}\nonumber
\end{equation}

\noindent that corresponds precisely to the limiting case $E_{\mathrm{sol}} = E_{\mathrm{tot}}$, and  provides an upper bound for the soliton mass for a fixed total energy.  In contrast, the ``1/2 relation'' also defined in \cite{Blum2025JCAP}

\begin{equation}
M_{\mathrm{sol}} \propto \left(\frac{|E_{\mathrm{tot}}|}{M}\right)^{1/2}\nonumber
\end{equation}

\noindent accounts for the presence of bound non-solitonic mass and yields a smaller soliton fraction. Repeating the above scaling argument using this relation leads, for equal-mass mergers, to a characteristic fraction of order

\begin{equation}
\frac{M_{c,{\rm final}}}{M_{c,1}+M_{c,2}} \sim 0.5 .\nonumber
\end{equation}

\noindent We therefore expect that for $g=0$ the final core mass fraction lies within the band

\begin{equation}
0.5 \lesssim 
\frac{M_{c,{\rm final}}}{M_{c,1}+M_{c,2}}
\lesssim 0.63 ,
\label{eq:band}
\end{equation}

\noindent where the upper limit corresponds to maximal condensation efficiency.

The nearly universal mass fraction in binary soliton mergers for $g=0$ is thus expected. It follows from the cubic energy scaling $E_{\mathrm{sol}} \propto -M^3$ of ground state solutions, energy conservation during the merger, gravitational cooling that removes positive energy, and the binding energy distributed between solitonic and non-solitonic components.

% ---------------------------------------------------------- subsection

\subsubsection{Generalization to $g\neq 0$}

The previous case is based on the cubic scaling of the energy with the mass
$E_{\mathrm{sol}} \propto -M^3$. When self-interaction is included, the structure of the equilibrium solution changes and so does the mass-energy scaling relation. For the GPP system, the total energy contains now three contributions,
$E = K + W + I$, where 
$K \sim \frac{M}{r_c^2}$, 
$W \sim -\frac{M^2}{r_c}$, 
$I \sim \frac{g M^2}{r_c^3}$.
Depending on which term balances gravity, different scaling relations between mass and radius should appear. 
In general, the soliton energy can be written as an effective power law

\begin{equation}
E_{\mathrm{sol}} \propto - M^a ,
\qquad a = a(g),
\nonumber
%\label{eq:general_alpha}
\end{equation}

\noindent where the exponent $a$ depends on the strength and sign of the self-interaction. Repeating the energetic argument of the previous case with this generalized scaling, the total initial energy of two separated cores is

\begin{equation}
E_{\rm initial} \sim -M_{c,1}^a - M_{c,2}^a .\nonumber
\end{equation}

\noindent Again, assuming maximal condensation possible, the final total energy is the energy of the soliton

\begin{equation}
E_{\mathrm{sol,final}} = E_{\mathrm{tot,final}},\nonumber
\end{equation}

\noindent and one obtains

\begin{equation}
M_{c,{\rm final}}^a \sim M_{c,1}^a + M_{c,2}^a .\nonumber
\end{equation}

\noindent For equal-mass mergers ($\lambda = 1$) $M_{c,1} = M_{c,2}=M_0$, this implies

\begin{equation}
M_{c,{\rm final}}^a = 2 M_0^a ~~ \Rightarrow ~~ 
M_{c,{\rm final}} = 2^{1/a} M_0,\nonumber
\end{equation}

\noindent and after dividing by the total initial core mass $2M_0$, the final-to-initial core mass fraction becomes
\begin{equation}
\frac{M_{c,{\rm final}}}{M_{c,1}+M_{c,2}}=2^{(1-a)/a}.
\label{eq:magic_general}
\end{equation}

\noindent This expression makes explicit that the magic mass fraction is controlled by the effective mass--energy exponent $a$ of the equilibrium configuration. Now, let us explore three different regimes and see whether the value of $a$ can be related to the physical mechanism supporting the core against gravity.

{\it Quantum-pressure dominated regime} ($g \approx 0$). As seen before, the quantum kinetic term balances gravity ($K \sim |W|$), and one recovers

\begin{equation}
r_c \propto M^{-1}, ~~~ E \propto -M^3 ,\nonumber
\end{equation}

\noindent so that $a = 3$ and

\begin{equation}
\frac{M_{c,{\rm final}}}{M_{c,1}+M_{c,2}} =    2^{-2/3} \simeq 0.63 .\nonumber
\end{equation}

\noindent as before.

{\it Thomas--Fermi regime} ($g>0$). If the self-interaction pressure dominates over quantum pressure ($I \sim |W|$), say for $g=g_{TF}$, one finds

\begin{equation}
r_c \sim \text{const},~~ E \propto -M^2 ,\nonumber
\end{equation}

\noindent so that $a = 2$ and

\begin{equation}
\frac{M_{c,{\rm final}}}{M_{c,1}+M_{c,2}}  =   2^{-1/2}  \simeq 0.707 .\nonumber
\end{equation}

\noindent In this regime the merger becomes more efficient at retaining mass in the core, and the expected fraction increases to $\sim 0.7$. In our simulations we do not consider such a big self-interaction dominated scenario, which is why the ratio found does not approach 0.707.

{\it Intermediate regime} ($0<g<g_{\rm TF}$). For moderate repulsive self-interaction both quantum pressure and self-interaction contribute to the equilibrium structure. In this case $a$ is expected to be within the range

\begin{equation}
2 < a < 3 ,\nonumber
\end{equation}

\noindent and the mass fraction should smoothly change between $0.63$ and $0.707$ according to Eq. \eqref{eq:magic_general}. Notice that this is precisely what we find in our simulations, as seen in Table \ref{tab:mass_core_ratio}.

{\it Attractive self-interaction} ($g<0$). For negative $g$, the self-interaction enhances gravitational binding and makes the core more compact. The effective mass-energy scaling becomes steeper

\begin{equation}
a > 3 ,
\end{equation}

\noindent which, through Eq. \eqref{eq:magic_general}, leads to a smaller final core fraction. In this case one expects

\begin{equation}
\frac{M_{c,{\rm final}}}{M_{c,1}+M_{c,2}} < 0.63 ,\nonumber
\end{equation}

\noindent which is consistent with the trend toward less efficient core retention from our simulations as also seen in Table \ref{tab:mass_core_ratio}.

In this way, the magic fraction is universal for each value of $g$, which changes from one value to another, and the transition from attractive to repulsive self-interaction also corresponds to a change in $a$, which implies a shift of the {\bf magic} mass fraction as found from our simulations.

%----------------------------------------------------------%
% ---------------->   Section     --------------------% 
% ------------------------------- -------------------------%
\section{Effects of the presence of a gas}
\label{sec:ideal-gas}

% -----------
\subsection{Equations}

The dynamics of the bosonic cloud coupled to the ideal gas dynamics 
of a compressible fluid is the Gross-Pitaevskii-Poisson-Euler (GPPE) 
system, which %after a rescaling from physical to code units as explained in Appendix \ref{app:unitsSEP} 
reads:

\begin{eqnarray}
    i\partial_{t} \Psi = \left(-\dfrac{1}{2}\nabla^2 + V + g|\Psi|^2\right)\Psi, \label{eq:Schrodinger_GPPE} \\
    \partial_{t} \rho + \nabla \cdot (\rho \vec{v}) = 0, \label{eq:Cons_Mass_GPPE} \\
    \partial_{t} (\rho \vec{v}) + \nabla \cdot (\rho \vec{v}\otimes\vec{v} + pI) = -\rho\nabla V, \label{eq:Cons_Momentum_GPPE} \\
    \partial_{t} E + \nabla \cdot [(E + p)\vec{v}] = -\rho\vec{v}\cdot\nabla V, \label{eq:Cons_Energy_GPPE} \\
    \nabla^2 V = \rho + |\Psi|^2, \label{eq:Poisson_GPPE}
\end{eqnarray}

\noindent where $\Psi$ and $|\Psi|^2$ are the order parameter and bosonic cloud density as before, whereas the fluid variables of a volume element are its density $\rho$, velocity $\vec{v}$, pressure $p$ and total energy $E = \rho (e + \frac{1}{2}|\vec{v}|^2)$, where $e$ is the specific internal energy, and $I$ represents the unit matrix. The gravitational potential $V$ is sourced by the contribution of the bosonic component and the fluid. To close the system, we use an equation of state that relates the pressure $p$, density $\rho$ and  internal energy $e$.  Since we want to explore the effects of the gas we consider the case with $g=0$.

% -----------
\subsection{Initial Conditions and Numerical Setup}

Our initial configurations are Newtonian Fermion Boson Stars (FBS), which 
are stationary solutions of the GPPE system 
(\ref{eq:Schrodinger_GPPE})-(\ref{eq:Poisson_GPPE}) above 
(see \cite{Alvarez_Rios_2023} for details of their construction), assuming the 
gas obeys a polytropic equation of state:

\begin{equation}
    p = K_p\rho^{1+1/n},
    \label{eq:polytropic}    
\end{equation}

\noindent where $K_p$ is the polytropic constant and $n$ the polytropic index.  Like in the case of BSs, these solutions are stable under spherical perturbations \cite{Alvarez_Rios_2023}, and in fact have attractor properties in a general scenario \cite{FermionBosoStars2024}. When evolved under an ideal gas equation of state,

\begin{equation}
    p = (\gamma - 1)\rho e,
    \label{eq:ideal_gas}    
\end{equation}

\noindent where $\gamma$ is the adiabatic index, an arbitrary initial random distribution of ideal gas approximately evolves toward an FBS \cite{FermionBosoStars2024}. This implies that the evolution of the ideal gas follows an adiabatic and isentropic process described by the polytropic equation of state \cite{Alvarez_Rios_2023}.

The FBS solutions are characterized by the central values $(\psi_c, \rho_c)$.
The pressure is defined according to the polytropic equation of state (\ref{eq:polytropic}) with $n = 1.5$ and $K_p = 10$. The polytropic index $n$ corresponds to an adiabatic index $\gamma = 1 + 1/n = 5/3$ for a monoatomic gas. The choice of the polytropic constant value does not affect the generality of the system due to the rescaling invariance of the GPPE equations \cite{Alvarez_Rios_2023}. This allows us to arbitrarily set the value of $K_p$, since any other choice can be recovered by scaling.

Also important to our analysis is the relative contribution of the gas to the FBS, defined by the mass ratio:

\begin{equation}
    MR = \frac{M_{\text{gas}}}{M_{\text{BEC}}},
\end{equation}

\noindent where $M_{\text{gas}}$ and $M_{\text{BEC}}$ are the total masses of the fluid and bosonic components, respectively.

\textit{The Binary FBS System.} The bosonic part of the FBS system is initialized in the same way as in the pure boson case, using equation (\ref{eq:initial_order_parameter_BS}), but now the wavefunctions \(\psi_{1}\) and \(\psi_2\) correspond to FBS solutions. The fluid component however, requires additional initialization:

\begin{equation}
    \rho(0, \vec{x}) = \max(\rho_1 + \rho_2, floor),
    \label{eq:initial_density_FBS}
\end{equation}

\noindent  where $floor = 10^{-10}$ is an artificial atmosphere introduced to prevent division by zero densities in Euler equations. The velocity field for the gas is defined as:

\begin{equation}
    \vec{v}(0, \vec{x}) = 
    \begin{cases}
    \vec{v}_1, & \text{if} \quad |\vec{x}-\vec{x}_1| < R_1, \\
    \vec{v}_2, & \text{if} \quad |\vec{x}-\vec{x}_2| < R_2, \\
    (0,0,0), & \text{otherwise},
    \end{cases}    
\end{equation}

\noindent where $R_1$ and $R_2$ are the radii of the individual FBS configurations, defined as the first zero of the polytropic density profile for each of them.

% -----------
\subsection{Numerical Scheme and Simulations Setup}

With the compressible fluid aside from the boson dynamics, 
we use the coupled version of CAFE-FDM \cite{AlvarezGuzman2022}, 
that solves the GPPE system using a third-order Runge-Kutta method 
for time integration. The Gross-Pitaevskii equation 
(\ref{eq:Schrodinger_GPPE}) and Poisson equation 
(\ref{eq:Poisson_GPPE}) are solved using the Fast Fourier Transform 
(FFT).  Euler equations 
(\ref{eq:Cons_Mass_GPPE})-(\ref{eq:Cons_Energy_GPPE}) are solved 
using high-resolution shock-capturing (HRSC) methods, specifically with 
the Minmod slope limiter and the HLLE approximate Riemann solver to 
compute the numerical fluxes.

The boundary conditions used are periodic for both the wave function and the hydrodynamic variables. This choice ensures the conservation of mass and total energy. Again, for the gravitational potential we use Dirichlet boundary conditions imposed by setting \( V = 0 \) at the boundaries of the domain. 

The evolution is carried out in the same spatial domain as before \( D = [-20, 20]^3 \), discretized using \( N = 128 \) grid points in each direction, corresponding to a spatial resolution of \( h = 0.3125 \). The system is evolved during 250 time units with resolution of \( \Delta t = 0.01 \), which satisfies the Courant condition \( \Delta t / h^2 < \dfrac{1}{6 \pi^2} \).

% -----------
\subsection{Results of simulations}

We remind that in this case, we consider only the contribution of the polytropic gas, for which we switch off self-interaction, $g =0$. 
Using again Fig. \ref{fig:binary_merger_illustration} to explain initial data,
the configuration on the left is defined with fixed 
parameters, specifically with a normalized central wave function 
amplitude of $\psi_{c,1} = 1$.
The initial velocity and position of the initial configurations  are the same as for Newtonian Boson Stars. 
The right-hand FBS in Fig. \ref{fig:binary_merger_illustration} is initialized 
with scale factor $\lambda = \sqrt{1.5}$, two values of the
initial impact parameter $y_2=5,10$, and five values of the initial 
velocity $v_{x,1}=$0.05, 0.1, 0.15, 0.2, 0.25, which is a set of ten 
simulations. We explore three values
of the mass ratios between the 
gas and BEC $MR = 0.1$, $1.0$ and $10.0$ which samples different 
dominance scenarios. Initial positions and 
velocity are chosen to ensure that the center of mass of the system 
remains near the origin.

\begin{figure}
    \centering
    \includegraphics[width=\linewidth]{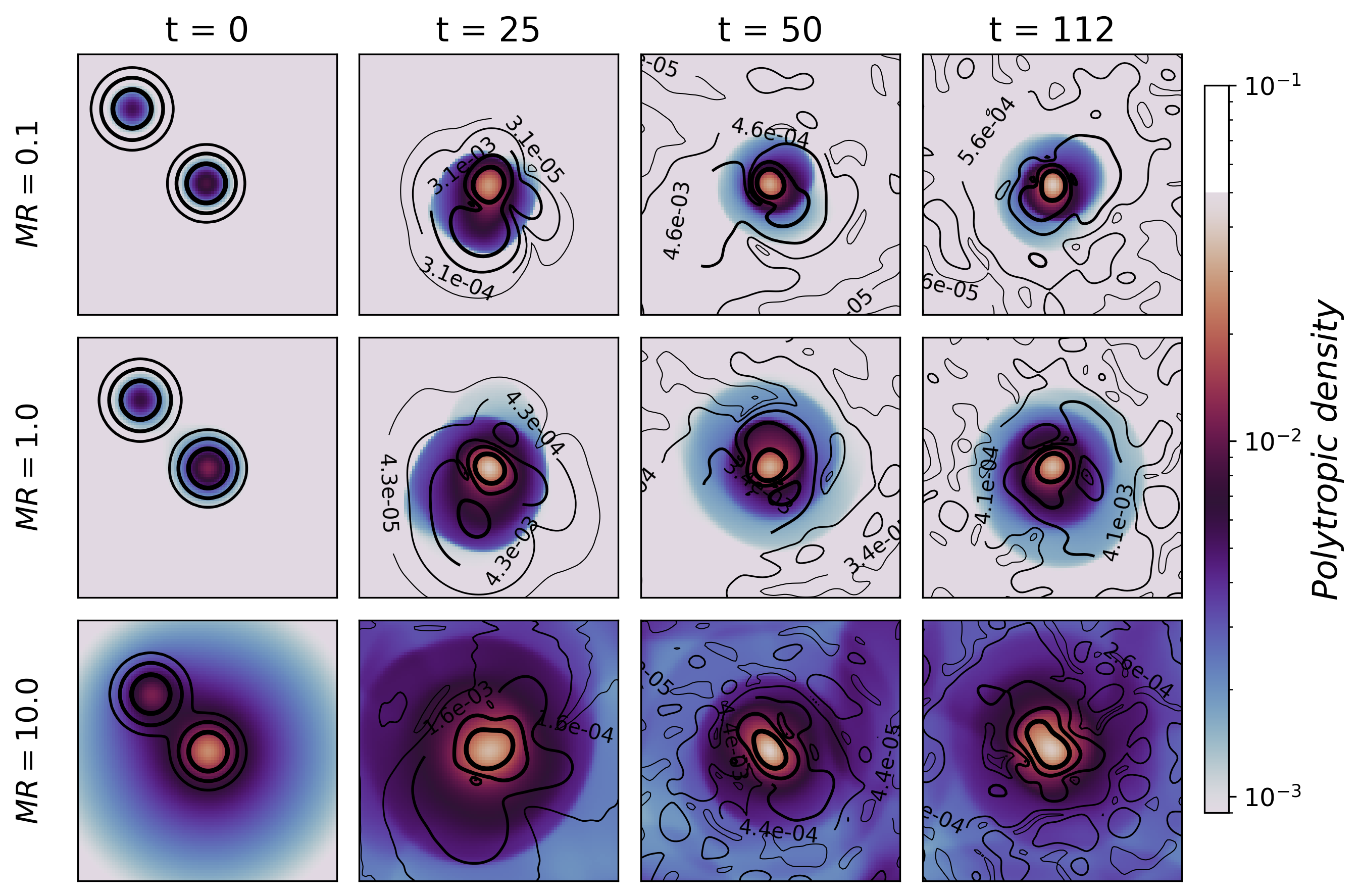}
    \caption{Snapshots of the density evolution on the plane \(z = 0\) at 
various  times (\(t = 0\), \(25\), \(50\), and \(112\)). Each row 
corresponds to a different value of the mass ratio \(MR = 0.1\), 1.0 
and 10.0. The bosonic and polytropic gas components are represented by 
isocontours and a color map, respectively. The merger leads to the 
formation of a new BEC soliton core at the center. The parameters used in these 
cases are \(\lambda = \sqrt{1.5}\), \(y_1 = 5\) and \(v_{x_1} = 0.2\).}
    \label{fig:evolutionFBS}
\end{figure}

\textit{Core Formation.} Figure~\ref{fig:evolutionFBS} shows snapshots 
of the density evolution on the \(z = 0\) plane at different times 
\(t = 0, 25, 50, 112\), for the three values of the mass ratio 
\(MR = 0.1, 1.0\) and \(10.0\) with the 
scale factor $\lambda = \sqrt{1.5}$, impact parameter $y_1 = 5$ and
 $v_{x_1} = 0.2$.

The color map represents the polytropic gas density, the  isocontours 
denote the bosonic cloud density, where thicker lines indicate higher 
density values. In all cases, the initial scale factor for the FBS is 
set to \(\lambda = \sqrt{1.5}\), and the same initial impact parameter 
and velocity (\(y_1 = 5\), \(v_{x_1} = 0.2\)) are used for the three 
values of $MR$. 

After the collision, the bosonic cloud forms a core surrounded by an 
extended halo of BEC, the gas component evolves depending on the value 
of \(MR\). For \(MR = 0.1\) and 1.0, the gas is subdominant and disperses 
quickly during the merger, with only a small fraction accumulating 
near the newly formed soliton core at late times. The resulting gas 
distribution is diffuse and does not strongly affect the core structure. 

In the case with \(MR = 10.0\), the gas dominates the mass contents
and disperses throughout the domain. Despite the 
gas dominance, the bosonic component merges and forms a core as in the 
previous cases, which develops a potential well that serves as an 
accumulation point for the gas.

\begin{figure}
    \centering
    \includegraphics[width=\linewidth]{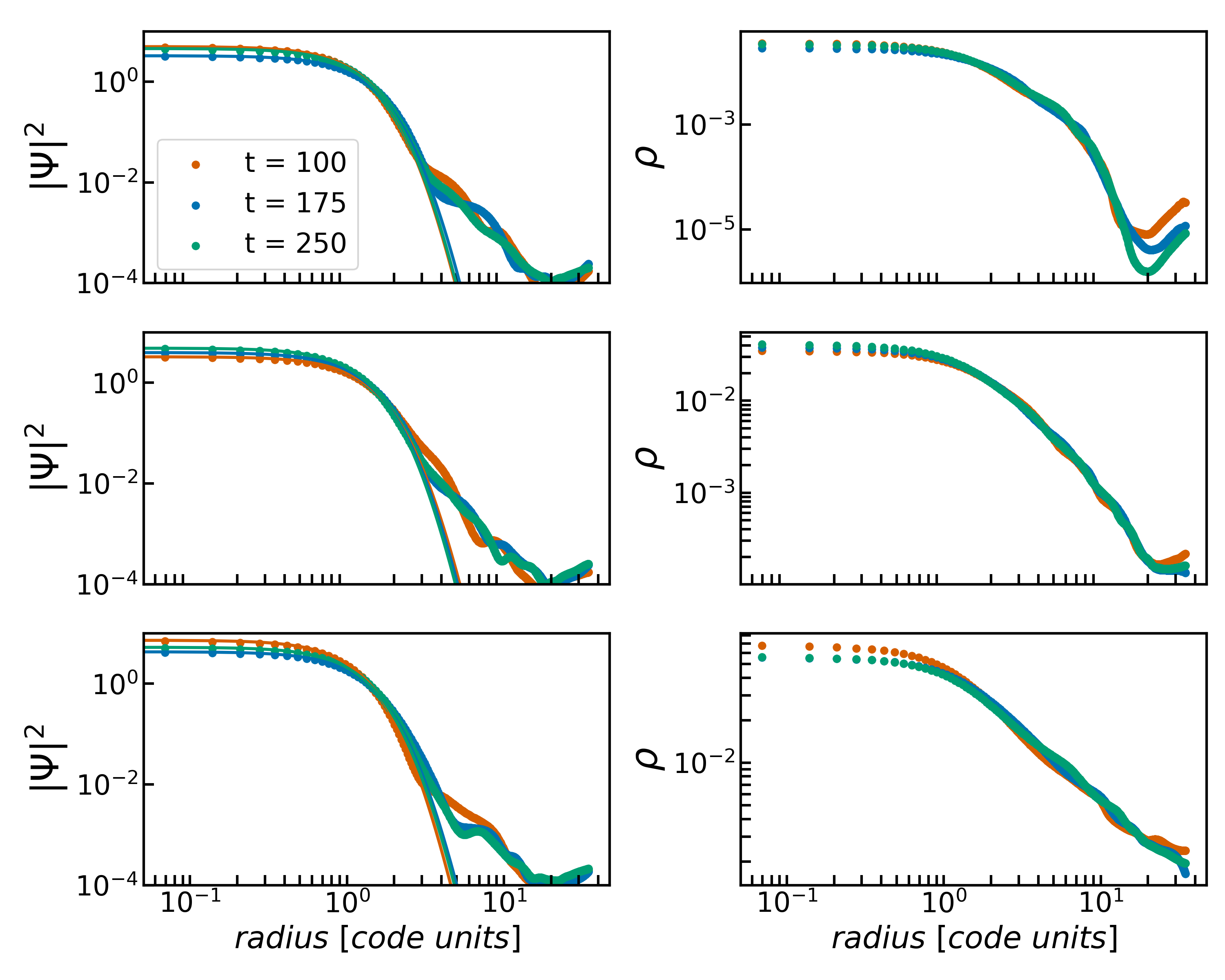}
    \caption{(Left) Angle-averaged density profile (\ref{eq:average_density}) at three different times and different values of \(MR = 0.1\), 1.0 and 10, from top to bottom, compared to the empirical fitting model (\ref{eq:soliton_density}) in solid lines.  (Right) The angle-averaged density profile (\ref{eq:average_density}) of the gas component for the same different times, indicating that the profile is nearly stationary. The simulation uses the parameters \(y_1 = 5\) and \(v_{x_1} = 0.2\).}
    \label{fig:avg_fit_FBS}
\end{figure}

Figure~\ref{fig:avg_fit_FBS} shows the angle-averaged density profiles 
of both the bosonic cloud and the polytrope at different times. The 
left-hand column compares the bosonic profile to the empirical solitonic 
fit that uses the core model Eq. (\ref{eq:soliton_density}), showing excellent 
agreement regardless of \(MR\). The right-hand column displays the 
evolution of the gas profile, showing that for small values of \(MR\), 
the gas is almost completely concentrated at the soliton, and 
as \(MR\) increases, the gas becomes disperse throughout the 
domain but still forms a core that evolves along with 
the final soliton core. 

Together, these results indicate a cooperative interaction where the 
bosons form a self-gravitating soliton, while the gas  responds to the 
evolving potential, leading to the formation of a 
long-living concentrated density of the merged structure. 
The final configuration is a bound core-halo system 
whose properties depend on the mass ratio \(MR\). 

\begin{figure}
    \centering
    \includegraphics[width=\linewidth]{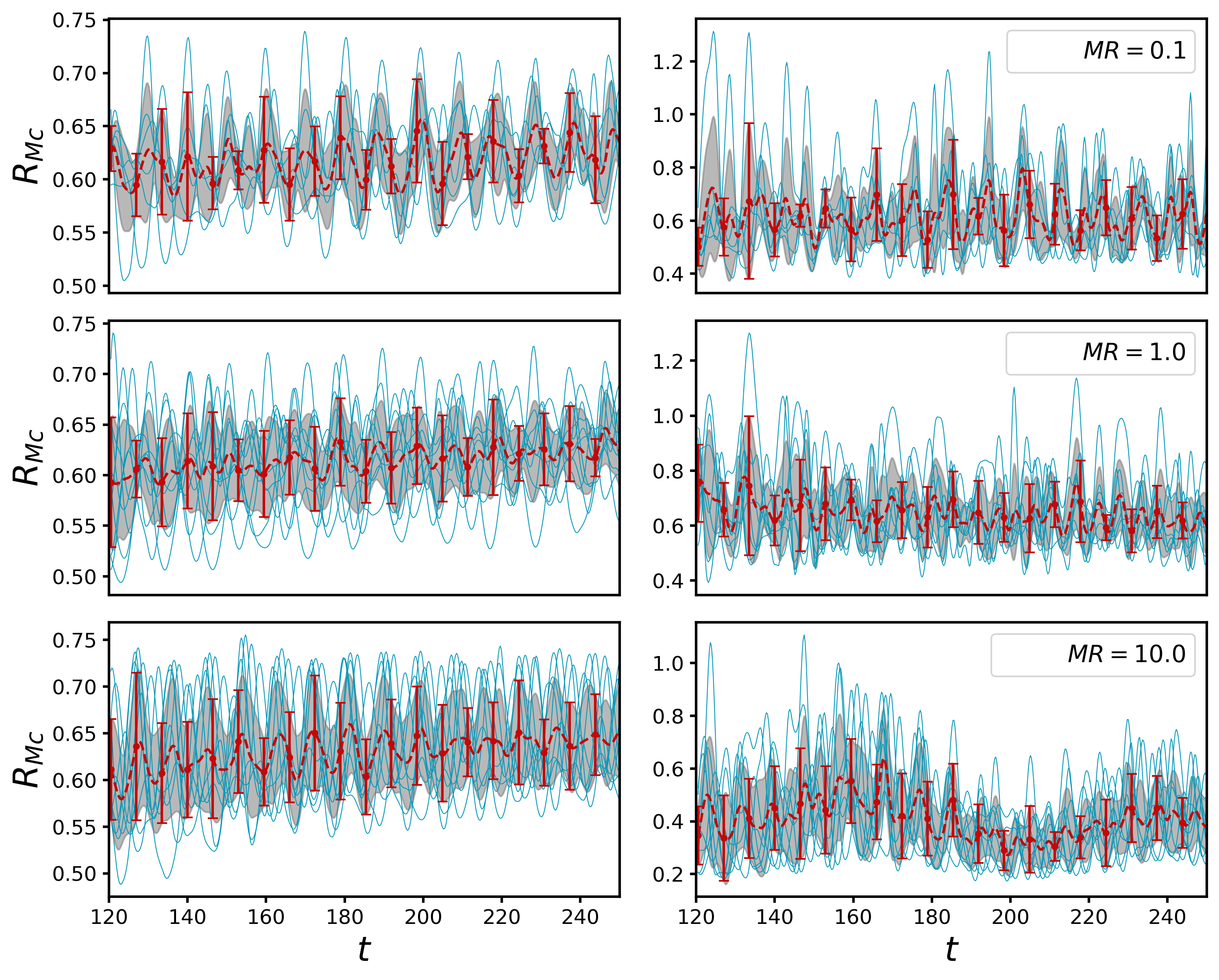}
    \caption{Evolution of the $R_{M_c}$ ratio for FBS mergers. The 
left column contains the \(R_{M_c}\) for the BEC component, while 
the right column  shows \(R_{M_c}\) for the polytropic gas. 
Each panel 
corresponds to 10 simulations for each fixed value of \(MR = 0.1\), 
1.0 and 10.0, from top to bottom, and different parameters \(y_1\) 
and \(v_{x_1}\). 
The thick dotted line represents the average value of the core mass 
ratio $R_{M_c}$, and the gray band indicates the standard deviation in time,
both calculated using the ten simulations for each value of $MR$. 
}
    \label{fig:mass_core_ratioFBS}
\end{figure}

\textit{Final to Initial Core Mass Ratio.}  We again calculate the ratio for the bosonic 
component as before using (\ref{eq:mass_core_ratio}) in order to know 
whether a universal value persists under the presence of the 
gas. On the other hand, there is no practical way to define the 
core of a polytrope, however in order to have a handy way to compare 
against the bosons we use the same recipe and define the core radius 
of the polytrope as that where the angular average density $f=\rho$ 
in Eq. (\ref{eq:average_density}) is half the central value. Figure 
\ref{fig:mass_core_ratioFBS} shows the evolution of the core mass 
ratio separately for the bosons and  for the gas, for all simulations. 
Each row corresponds to a fixed value of the mass ratio \(MR = 0.1\), 
1.0 and 10.0 from top to bottom. The data shows that \(R_{M_c}\) 
for the bosonic part fluctuates around a well-defined average for the 
three values of $MR$.

\begin{table}
   \centering
   Bosons\\
   \hspace{1cm}\\
   \begin{tabular}{|c|c|c|c|c|c|}
   \hline
   \(MR\) & 0.1 & 1.0 & 10.0 \\
   \hline
   $\langle R_{M_c,\text{avg}} \rangle$ & 0.617 & 0.615 & 0.630  \\
   \hline
   $\langle \sigma\rangle$ & 0.034  & 0.038 & 0.049  \\
   \hline
   \end{tabular}\\
\hspace{1cm}\\
%%%%
%   \centering
   \hspace{1cm}\\
   Gas\\
   \hspace{1cm}\\
   \begin{tabular}{|c|c|c|c|c|c|}
   \hline
   \(MR\) & 0.1 & 1.0 & 10.0 \\
   \hline
   $\langle R_{M_c,\text{avg}}\rangle$ & 0.614 & 0.646 & 0.407  \\
   \hline
   $\langle\sigma\rangle$ & 0.121 & 0.111 & 0.130 \\
   \hline
   \end{tabular}
%%%%
   \caption{Average in time of $R_{M_c}$ of the plots in Fig. 
\ref{fig:mass_core_ratioFBS} for the bosonic part, as well as 
the average of the standard deviation in time.}
   \label{tab:RMc_FBS_BECDM}
\end{table}

Table ~\ref{tab:RMc_FBS_BECDM} summarizes the 
average (\(R_{M_c,\text{avg}}\)) for the bosonic and polytropic component 
along with the standard deviation. We notice that for bosons the
$\sim 0.62$ factor remains despite the value of the dominance 
$MR$, whereas the polytropic gas does not, first because $\sigma$ is 
of order 20\% and second because when the polytrope dominates, the
value of $R_{M_c,\text{avg}}$ decreases with $MR$. This indicates
that binary systems in which the polytropic gas dominates are less 
able to compact the matter in the central core during their evolution. 
%The gas pressure disperses the whole system away.

% ------------------------------------------------------- subsection
\subsection{Explanation}

The persistence of the value $R_{M_c}\simeq 0.62$ for the bosonic component independently of the mass ratio $MR$, can be understood from energetic scaling arguments applied to the bosonic sector. In the present case we use $g=0$,  thus the bosonic dynamics is governed by the standard balance between quantum kinetic energy and self-gravity. For a bosonic core of mass we have the equilibrium energy scaling $E_{\text{bosons}} \propto -M_{\text{bosons}}^3$, which leads to the nearly universal merger fraction $R_{M_c} \approx 2^{-2/3}$ found in the pure boson case.

When the polytropic gas is included, the only coupling between the components is gravitational, through the common potential $V$ in Eq. (\ref{eq:Poisson_GPPE}). The mixed gravitational term scales as $W_{\text{int}} \sim - M_{\text{bosons}} M_{\text{gas}}/R$, where $R$ characterizes the size of the configuration. However, in our simulations the gas distribution remains more extended than the bosonic core for all explored values of $MR$. In that regime, the gravitational potential generated by the gas varies slowly across the bosonic core, and its contribution to the energy is approximately linear in $M_{\text{bosons}}$ with negligible dependence on $r_c$. Thus, the mass--radius relation of the bosonic part and its associated cubic energy scaling remain essentially unchanged.

As a consequence, during the merger, the bosonic component redistributes its mass according to the same energetic principle as in the pure case, leading to $R_{M_c}\simeq 0.62$ for all values of $MR$ reported in Table~\ref{tab:RMc_FBS_BECDM}. In contrast, the gas component does not obey a universal mass--radius relation of the same type. Its equilibrium is determined by the balance between gravity and pressure gradients, and its total energy scaling depends explicitly on the polytropic index. Moreover, hydrodynamic effects such as shocks and pressure-driven expansion introduce additional channels for energy redistribution. For this reason, no universal merger fraction is expected for the gas, and the larger dispersion and decrease of $R_{M_c}$ for high $MR$ reflect the role of pressure in preventing a clear central concentration of gas density.

In summary, the simulations indicate that the universality of the bosonic core mass fraction is a property of the bosonic sector, which preserves its cubic energy scaling even in the presence of a dominant but spatially extended gas component. The gas only reshapes the global gravitational background, while the formation and mass fraction of the solitonic BEC core remain controlled by the intrinsic energetics of the bosonic part.

%----------------------------------------------------------%
% ---------------->   Section     --------------------% 
% ------------------------------- -------------------------%
\section{Conclusions}
\label{sec:conclusions}

In this work, we have performed a systematic numerical study of the merger 
of dark matter cores that include self-interaction and a coupled baryonic 
component modeled as an ideal gas. 
Our results allow us to draw some conclusions on the effects of 
self-interaction and gas dynamics in the merger of two equilibrium 
configurations.

The introduction of the self-interaction parameter $g$ generalizes the finding
of a constant value of $R_{M_c}$ in \cite{Schwabe:2016}, and implies that repulsive 
interactions ($g > 0$) act as an additional pressure term that stabilizes 
the core against tidal disruption during the merger, resulting in a 
systematic increase of $R_{M_c}$. Conversely, attractive interactions 
($g < 0$) promote a more violent relaxation and higher mass ejection, 
leading to smaller remnants. 
This shift indicates that the final mass of solitonic cores is 
sensitive to the fundamental properties of the dark matter particle, 
with potential implications for the core-halo mass relation.
The numerical trends reported above can be understood from simple energetic scaling arguments.

In the case with an ideal gas component and $g=0$, the coupling between the bosonic and gas sectors is purely gravitational. Because the gas remains more spatially extended than the solitonic core, its contribution to the gravitational potential acts as a slowly varying background that does not modify the bosonic mass--radius relation. During the merger, we find that the bosonic component continues to drive the condensation process and forms the final solitonic core, while the gas settles into a diffuse distribution that traces the gravitational potential well created by the bosonic soliton. Even in cases where the gas dominates the total mass ($MR=10.0$), no compact gaseous core forms. As a consequence, the cubic energy scaling of the bosonic component and the corresponding nearly universal value of $R_{M_c}$ are preserved. These results explain why the solitonic merger fraction remains robust even in the presence of a significant gas component.

In summary, while the fundamental nature of the BEC DM core as a merger remnant remains unchanged, its quantitative properties, such as total mass and retention ratio, are significantly modulated by its environment and internal interactions. These effects should be considered when calibrating semi-analytical models of galaxy formation and when interpreting the central density profiles of dark matter halos.

% ----->     ACKNOWLEDGMENTS     <-----
\section*{Acknowledgments}
This research is supported by grants CIC-UMSNH-4.9 and  SECIHTI Grant No. CFB-2025-I-759. 

\section*{Data availability}
 The data that support the findings of this article are openly available \cite{ourdata}.

%PRS acknowledges support from NASA under Grant No. 80NSSC22K175, and thanks Taha Dawoodbhoy and Luis Padilla for discussion. 
% -------------------------------------------------------
% -----     REFERENCES     ----------
% -------------------------------------------------------
\bibliography{BECDM}

\begin{thebibliography}{32}%
\makeatletter
\providecommand \@ifxundefined [1]{%
 \@ifx{#1\undefined}
}%
\providecommand \@ifnum [1]{%
 \ifnum #1\expandafter \@firstoftwo
 \else \expandafter \@secondoftwo
 \fi
}%
\providecommand \@ifx [1]{%
 \ifx #1\expandafter \@firstoftwo
 \else \expandafter \@secondoftwo
 \fi
}%
\providecommand \natexlab [1]{#1}%
\providecommand \enquote  [1]{``#1''}%
\providecommand \bibnamefont  [1]{#1}%
\providecommand \bibfnamefont [1]{#1}%
\providecommand \citenamefont [1]{#1}%
\providecommand \href@noop [0]{\@secondoftwo}%
\providecommand \href [0]{\begingroup \@sanitize@url \@href}%
\providecommand \@href[1]{\@@startlink{#1}\@@href}%
\providecommand \@@href[1]{\endgroup#1\@@endlink}%
\providecommand \@sanitize@url [0]{\catcode `\\12\catcode `\$12\catcode
  `\&12\catcode `\#12\catcode `\^12\catcode `\_12\catcode `\%12\relax}%
\providecommand \@@startlink[1]{}%
\providecommand \@@endlink[0]{}%
\providecommand \url  [0]{\begingroup\@sanitize@url \@url }%
\providecommand \@url [1]{\endgroup\@href {#1}{\urlprefix }}%
\providecommand \urlprefix  [0]{URL }%
\providecommand \Eprint [0]{\href }%
\providecommand \doibase [0]{http://dx.doi.org/}%
\providecommand \selectlanguage [0]{\@gobble}%
\providecommand \bibinfo  [0]{\@secondoftwo}%
\providecommand \bibfield  [0]{\@secondoftwo}%
\providecommand \translation [1]{[#1]}%
\providecommand \BibitemOpen [0]{}%
\providecommand \bibitemStop [0]{}%
\providecommand \bibitemNoStop [0]{.\EOS\space}%
\providecommand \EOS [0]{\spacefactor3000\relax}%
\providecommand \BibitemShut  [1]{\csname bibitem#1\endcsname}%
\let\auto@bib@innerbib\@empty
%</preamble>
\bibitem [{\citenamefont {Matos}\ and\ \citenamefont
  {Urena-Lopez}(2001)}]{Matos:2000ss}%
  \BibitemOpen
  \bibfield  {author} {\bibinfo {author} {\bibfnamefont {Tonatiuh}\
  \bibnamefont {Matos}}\ and\ \bibinfo {author} {\bibfnamefont {L.~Arturo}\
  \bibnamefont {Urena-Lopez}},\ }\bibfield  {title} {\enquote {\bibinfo {title}
  {{A Further analysis of a cosmological model of quintessence and scalar dark
  matter}},}\ }\href {\doibase 10.1103/PhysRevD.63.063506} {\bibfield
  {journal} {\bibinfo  {journal} {Phys. Rev. D}\ }\textbf {\bibinfo {volume}
  {63}},\ \bibinfo {pages} {063506} (\bibinfo {year} {2001})}\BibitemShut
  {NoStop}%
%%CITATION = ASTRO-PH/0006024;%%
\bibitem [{\citenamefont {Hu}\ \emph {et~al.}(2000)\citenamefont {Hu},
  \citenamefont {Barkana},\ and\ \citenamefont {Gruzinov}}]{Hu:2000ke}%
  \BibitemOpen
  \bibfield  {author} {\bibinfo {author} {\bibfnamefont {Wayne}\ \bibnamefont
  {Hu}}, \bibinfo {author} {\bibfnamefont {Rennan}\ \bibnamefont {Barkana}}, \
  and\ \bibinfo {author} {\bibfnamefont {Andrei}\ \bibnamefont {Gruzinov}},\
  }\bibfield  {title} {\enquote {\bibinfo {title} {{Cold and fuzzy dark
  matter}},}\ }\href {\doibase 10.1103/PhysRevLett.85.1158} {\bibfield
  {journal} {\bibinfo  {journal} {Phys. Rev. Lett.}\ }\textbf {\bibinfo
  {volume} {85}},\ \bibinfo {pages} {1158--1161} (\bibinfo {year}
  {2000})}\BibitemShut {NoStop}%
%%CITATION = ASTRO-PH/0003365;%%
\bibitem [{\citenamefont {Hui}\ \emph {et~al.}(2017)\citenamefont {Hui},
  \citenamefont {Ostriker}, \citenamefont {Tremaine},\ and\ \citenamefont
  {Witten}}]{Hui:2016}%
  \BibitemOpen
  \bibfield  {author} {\bibinfo {author} {\bibfnamefont {Lam}\ \bibnamefont
  {Hui}}, \bibinfo {author} {\bibfnamefont {Jeremiah~P.}\ \bibnamefont
  {Ostriker}}, \bibinfo {author} {\bibfnamefont {Scott}\ \bibnamefont
  {Tremaine}}, \ and\ \bibinfo {author} {\bibfnamefont {Edward}\ \bibnamefont
  {Witten}},\ }\bibfield  {title} {\enquote {\bibinfo {title} {Ultralight
  scalars as cosmological dark matter},}\ }\href {\doibase
  10.1103/PhysRevD.95.043541} {\bibfield  {journal} {\bibinfo  {journal} {Phys.
  Rev. D}\ }\textbf {\bibinfo {volume} {95}},\ \bibinfo {pages} {043541}
  (\bibinfo {year} {2017})}\BibitemShut {NoStop}%
\bibitem [{\citenamefont {Schive}\ \emph {et~al.}(2014)\citenamefont {Schive},
  \citenamefont {Chiueh},\ and\ \citenamefont {Broadhurst}}]{Schive:2014dra}%
  \BibitemOpen
  \bibfield  {author} {\bibinfo {author} {\bibfnamefont {Hsi-Yu}\ \bibnamefont
  {Schive}}, \bibinfo {author} {\bibfnamefont {Tzihong}\ \bibnamefont
  {Chiueh}}, \ and\ \bibinfo {author} {\bibfnamefont {Tom}\ \bibnamefont
  {Broadhurst}},\ }\bibfield  {title} {\enquote {\bibinfo {title} {{Cosmic
  Structure as the Quantum Interference of a Coherent Dark Wave}},}\ }\href
  {\doibase 10.1038/nphys2996} {\bibfield  {journal} {\bibinfo  {journal}
  {Nature Phys.}\ }\textbf {\bibinfo {volume} {10}},\ \bibinfo {pages}
  {496--499} (\bibinfo {year} {2014})},\ \Eprint
  {http://arxiv.org/abs/1406.6586} {arXiv:1406.6586 [astro-ph.GA]} \BibitemShut
  {NoStop}%
\bibitem [{\citenamefont {Mocz}\ \emph {et~al.}(2017)\citenamefont {Mocz},
  \citenamefont {Vogelsberger}, \citenamefont {Robles}, \citenamefont {Zavala},
  \citenamefont {Boylan-Kolchin}, \citenamefont {Fialkov},\ and\ \citenamefont
  {Hernquist}}]{Mocz:2017wlg}%
  \BibitemOpen
  \bibfield  {author} {\bibinfo {author} {\bibfnamefont {Philip}\ \bibnamefont
  {Mocz}}, \bibinfo {author} {\bibfnamefont {Mark}\ \bibnamefont
  {Vogelsberger}}, \bibinfo {author} {\bibfnamefont {Victor~H.}\ \bibnamefont
  {Robles}}, \bibinfo {author} {\bibfnamefont {Jesus}\ \bibnamefont {Zavala}},
  \bibinfo {author} {\bibfnamefont {Michael}\ \bibnamefont {Boylan-Kolchin}},
  \bibinfo {author} {\bibfnamefont {Anastasia}\ \bibnamefont {Fialkov}}, \ and\
  \bibinfo {author} {\bibfnamefont {Lars}\ \bibnamefont {Hernquist}},\
  }\bibfield  {title} {\enquote {\bibinfo {title} {{Galaxy formation with BECDM
  I. Turbulence and relaxation of idealized haloes}},}\ }\href {\doibase
  10.1093/mnras/stx1887} {\bibfield  {journal} {\bibinfo  {journal} {Mon. Not.
  Roy. Astron. Soc.}\ }\textbf {\bibinfo {volume} {471}},\ \bibinfo {pages}
  {4559--4570} (\bibinfo {year} {2017})},\ \Eprint
  {http://arxiv.org/abs/1705.05845} {arXiv:1705.05845 [astro-ph.CO]}
  \BibitemShut {NoStop}%
%%CITATION = ARXIV:1705.05845;%%
\bibitem [{\citenamefont {Veltmaat}\ \emph {et~al.}(2018)\citenamefont
  {Veltmaat}, \citenamefont {Niemeyer},\ and\ \citenamefont
  {Schwabe}}]{Veltmaat_2018}%
  \BibitemOpen
  \bibfield  {author} {\bibinfo {author} {\bibfnamefont {Jan}\ \bibnamefont
  {Veltmaat}}, \bibinfo {author} {\bibfnamefont {Jens~C.}\ \bibnamefont
  {Niemeyer}}, \ and\ \bibinfo {author} {\bibfnamefont {Bodo}\ \bibnamefont
  {Schwabe}},\ }\bibfield  {title} {\enquote {\bibinfo {title} {Formation and
  structure of ultralight bosonic dark matter halos},}\ }\href {\doibase
  10.1103/physrevd.98.043509} {\bibfield  {journal} {\bibinfo  {journal}
  {Physical Review D}\ }\textbf {\bibinfo {volume} {98}} (\bibinfo {year}
  {2018}),\ 10.1103/physrevd.98.043509}\BibitemShut {NoStop}%
\bibitem [{\citenamefont {Desjacques}\ \emph {et~al.}(2018)\citenamefont
  {Desjacques}, \citenamefont {Kehagias},\ and\ \citenamefont
  {Riotto}}]{Desjacques:2017fmf}%
  \BibitemOpen
  \bibfield  {author} {\bibinfo {author} {\bibfnamefont {Vincent}\ \bibnamefont
  {Desjacques}}, \bibinfo {author} {\bibfnamefont {Alex}\ \bibnamefont
  {Kehagias}}, \ and\ \bibinfo {author} {\bibfnamefont {Antonio}\ \bibnamefont
  {Riotto}},\ }\bibfield  {title} {\enquote {\bibinfo {title} {{Impact of
  ultralight axion self-interactions on the large scale structure of the
  Universe}},}\ }\href {\doibase 10.1103/PhysRevD.97.023529} {\bibfield
  {journal} {\bibinfo  {journal} {Phys. Rev. D}\ }\textbf {\bibinfo {volume}
  {97}},\ \bibinfo {pages} {023529} (\bibinfo {year} {2018})},\ \Eprint
  {http://arxiv.org/abs/1709.07946} {arXiv:1709.07946 [astro-ph.CO]}
  \BibitemShut {NoStop}%
\bibitem [{\citenamefont {Schwabe}\ and\ \citenamefont
  {Niemeyer}(2022)}]{Gotinga2022}%
  \BibitemOpen
  \bibfield  {author} {\bibinfo {author} {\bibfnamefont {Bodo}\ \bibnamefont
  {Schwabe}}\ and\ \bibinfo {author} {\bibfnamefont {Jens~C.}\ \bibnamefont
  {Niemeyer}},\ }\bibfield  {title} {\enquote {\bibinfo {title} {Deep zoom-in
  simulation of a fuzzy dark matter galactic halo},}\ }\href {\doibase
  10.1103/PhysRevLett.128.181301} {\bibfield  {journal} {\bibinfo  {journal}
  {Phys. Rev. Lett.}\ }\textbf {\bibinfo {volume} {128}},\ \bibinfo {pages}
  {181301} (\bibinfo {year} {2022})}\BibitemShut {NoStop}%
\bibitem [{\citenamefont {Guzm\'an}\ and\ \citenamefont {Ure\~na
  L\'opez}(2004)}]{GuzmanUrena2004}%
  \BibitemOpen
  \bibfield  {author} {\bibinfo {author} {\bibfnamefont {F.~S.}\ \bibnamefont
  {Guzm\'an}}\ and\ \bibinfo {author} {\bibfnamefont {L.~Arturo}\ \bibnamefont
  {Ure\~na L\'opez}},\ }\bibfield  {title} {\enquote {\bibinfo {title}
  {Evolution of the schr\"odinger-newton system for a self-gravitating scalar
  field},}\ }\href {\doibase 10.1103/PhysRevD.69.124033} {\bibfield  {journal}
  {\bibinfo  {journal} {Phys. Rev. D}\ }\textbf {\bibinfo {volume} {69}},\
  \bibinfo {pages} {124033} (\bibinfo {year} {2004})}\BibitemShut {NoStop}%
\bibitem [{\citenamefont {Bernal}\ and\ \citenamefont
  {Guzm\'an}(2006{\natexlab{a}})}]{BernalGuzman2006b}%
  \BibitemOpen
  \bibfield  {author} {\bibinfo {author} {\bibfnamefont {Argelia}\ \bibnamefont
  {Bernal}}\ and\ \bibinfo {author} {\bibfnamefont {F.~S.}\ \bibnamefont
  {Guzm\'an}},\ }\bibfield  {title} {\enquote {\bibinfo {title} {Scalar field
  dark matter: Nonspherical collapse and late-time behavior},}\ }\href
  {\doibase 10.1103/physrevd.74.063504} {\bibfield  {journal} {\bibinfo
  {journal} {Physical Review D}\ }\textbf {\bibinfo {volume} {74}} (\bibinfo
  {year} {2006}{\natexlab{a}}),\ 10.1103/physrevd.74.063504}\BibitemShut
  {NoStop}%
\bibitem [{\citenamefont {Bernal}\ and\ \citenamefont
  {Guzm\'an}(2006{\natexlab{b}})}]{BernalGuzman2006a}%
  \BibitemOpen
  \bibfield  {author} {\bibinfo {author} {\bibfnamefont {Argelia}\ \bibnamefont
  {Bernal}}\ and\ \bibinfo {author} {\bibfnamefont {F.~S.}\ \bibnamefont
  {Guzm\'an}},\ }\bibfield  {title} {\enquote {\bibinfo {title} {Scalar field
  dark matter: Head-on interaction between two structures},}\ }\href {\doibase
  10.1103/physrevd.74.103002} {\bibfield  {journal} {\bibinfo  {journal}
  {Physical Review D}\ }\textbf {\bibinfo {volume} {74}} (\bibinfo {year}
  {2006}{\natexlab{b}}),\ 10.1103/physrevd.74.103002}\BibitemShut {NoStop}%
\bibitem [{\citenamefont {Paredes}\ and\ \citenamefont
  {Michinel}(2016)}]{PAREDES201650}%
  \BibitemOpen
  \bibfield  {author} {\bibinfo {author} {\bibfnamefont {Angel}\ \bibnamefont
  {Paredes}}\ and\ \bibinfo {author} {\bibfnamefont {Humberto}\ \bibnamefont
  {Michinel}},\ }\bibfield  {title} {\enquote {\bibinfo {title} {Interference
  of dark matter solitons and galactic offsets},}\ }\href {\doibase
  https://doi.org/10.1016/j.dark.2016.02.003} {\bibfield  {journal} {\bibinfo
  {journal} {Physics of the Dark Universe}\ }\textbf {\bibinfo {volume} {12}},\
  \bibinfo {pages} {50--55} (\bibinfo {year} {2016})}\BibitemShut {NoStop}%
\bibitem [{\citenamefont {{Schwabe}}\ \emph {et~al.}(2016)\citenamefont
  {{Schwabe}}, \citenamefont {{Niemeyer}},\ and\ \citenamefont
  {{Engels}}}]{Schwabe:2016}%
  \BibitemOpen
  \bibfield  {author} {\bibinfo {author} {\bibfnamefont {Bodo}\ \bibnamefont
  {{Schwabe}}}, \bibinfo {author} {\bibfnamefont {Jens~C.}\ \bibnamefont
  {{Niemeyer}}}, \ and\ \bibinfo {author} {\bibfnamefont {Jan~F.}\ \bibnamefont
  {{Engels}}},\ }\bibfield  {title} {\enquote {\bibinfo {title} {{Simulations
  of solitonic core mergers in ultralight axion dark matter cosmologies}},}\
  }\href {\doibase 10.1103/PhysRevD.94.043513} {\bibfield  {journal} {\bibinfo
  {journal} {\prd}\ }\textbf {\bibinfo {volume} {94}},\ \bibinfo {eid} {043513}
  (\bibinfo {year} {2016})},\ \Eprint {http://arxiv.org/abs/1606.05151}
  {arXiv:1606.05151 [astro-ph.CO]} \BibitemShut {NoStop}%
\bibitem [{\citenamefont {Seidel}\ and\ \citenamefont
  {Suen}(1994)}]{SeidelSuenCooling}%
  \BibitemOpen
  \bibfield  {author} {\bibinfo {author} {\bibfnamefont {Edward}\ \bibnamefont
  {Seidel}}\ and\ \bibinfo {author} {\bibfnamefont {Wai-Mo}\ \bibnamefont
  {Suen}},\ }\bibfield  {title} {\enquote {\bibinfo {title} {Formation of
  solitonic stars through gravitational cooling},}\ }\href {\doibase
  10.1103/PhysRevLett.72.2516} {\bibfield  {journal} {\bibinfo  {journal}
  {Phys. Rev. Lett.}\ }\textbf {\bibinfo {volume} {72}},\ \bibinfo {pages}
  {2516--2519} (\bibinfo {year} {1994})}\BibitemShut {NoStop}%
\bibitem [{\citenamefont {Guzm\'an}\ and\ \citenamefont {Ure\~na
  L\'opez}(2006)}]{GuzmanUrena2006}%
  \BibitemOpen
  \bibfield  {author} {\bibinfo {author} {\bibfnamefont {F.~S.}\ \bibnamefont
  {Guzm\'an}}\ and\ \bibinfo {author} {\bibfnamefont {L.~Arturo}\ \bibnamefont
  {Ure\~na L\'opez}},\ }\bibfield  {title} {\enquote {\bibinfo {title}
  {Gravitational cooling of self-gravitating bose condensates},}\ }\href
  {\doibase 10.1086/504508} {\bibfield  {journal} {\bibinfo  {journal} {The
  Astrophysical Journal}\ }\textbf {\bibinfo {volume} {645}},\ \bibinfo {pages}
  {814--819} (\bibinfo {year} {2006})}\BibitemShut {NoStop}%
\bibitem [{\citenamefont {\'Alvarez-Rios}\ \emph {et~al.}(2023)\citenamefont
  {\'Alvarez-Rios}, \citenamefont {Guzm\'an},\ and\ \citenamefont
  {Shapiro}}]{periodicas}%
  \BibitemOpen
  \bibfield  {author} {\bibinfo {author} {\bibfnamefont {Iv\'an}\ \bibnamefont
  {\'Alvarez-Rios}}, \bibinfo {author} {\bibfnamefont {Francisco~S.}\
  \bibnamefont {Guzm\'an}}, \ and\ \bibinfo {author} {\bibfnamefont {Paul~R.}\
  \bibnamefont {Shapiro}},\ }\bibfield  {title} {\enquote {\bibinfo {title}
  {Effect of boundary conditions on structure formation in fuzzy dark
  matter},}\ }\href {\doibase 10.1103/PhysRevD.107.123524} {\bibfield
  {journal} {\bibinfo  {journal} {Phys. Rev. D}\ }\textbf {\bibinfo {volume}
  {107}},\ \bibinfo {pages} {123524} (\bibinfo {year} {2023})}\BibitemShut
  {NoStop}%
\bibitem [{\citenamefont {Glennon}\ and\ \citenamefont
  {Prescod-Weinstein}(2021)}]{PhysRevD.104.083532}%
  \BibitemOpen
  \bibfield  {author} {\bibinfo {author} {\bibfnamefont {Noah}\ \bibnamefont
  {Glennon}}\ and\ \bibinfo {author} {\bibfnamefont {Chanda}\ \bibnamefont
  {Prescod-Weinstein}},\ }\bibfield  {title} {\enquote {\bibinfo {title}
  {Modifying pyultralight to model scalar dark matter with
  self-interactions},}\ }\href {\doibase 10.1103/PhysRevD.104.083532}
  {\bibfield  {journal} {\bibinfo  {journal} {Phys. Rev. D}\ }\textbf {\bibinfo
  {volume} {104}},\ \bibinfo {pages} {083532} (\bibinfo {year}
  {2021})}\BibitemShut {NoStop}%
\bibitem [{\citenamefont {Stallovits}\ and\ \citenamefont
  {Rindler-Daller}(2025)}]{TanjaBinary}%
  \BibitemOpen
  \bibfield  {author} {\bibinfo {author} {\bibfnamefont {Matthias}\
  \bibnamefont {Stallovits}}\ and\ \bibinfo {author} {\bibfnamefont {Tanja}\
  \bibnamefont {Rindler-Daller}},\ }\bibfield  {title} {\enquote {\bibinfo
  {title} {Single and merger soliton dynamics in scalar field dark matter with
  and without self-interactions},}\ }\href {\doibase
  10.1103/PhysRevD.111.023046} {\bibfield  {journal} {\bibinfo  {journal}
  {Phys. Rev. D}\ }\textbf {\bibinfo {volume} {111}},\ \bibinfo {pages}
  {023046} (\bibinfo {year} {2025})}\BibitemShut {NoStop}%
\bibitem [{\citenamefont {Su\'arez}\ and\ \citenamefont
  {Chavanis}(2017)}]{Suarez:2016eez}%
  \BibitemOpen
  \bibfield  {author} {\bibinfo {author} {\bibfnamefont {Abril}\ \bibnamefont
  {Su\'arez}}\ and\ \bibinfo {author} {\bibfnamefont {Pierre-Henri}\
  \bibnamefont {Chavanis}},\ }\bibfield  {title} {\enquote {\bibinfo {title}
  {{Cosmological evolution of a complex scalar field with repulsive or
  attractive self-interaction}},}\ }\href {\doibase 10.1103/PhysRevD.95.063515}
  {\bibfield  {journal} {\bibinfo  {journal} {Phys. Rev. D}\ }\textbf {\bibinfo
  {volume} {95}},\ \bibinfo {pages} {063515} (\bibinfo {year} {2017})},\
  \Eprint {http://arxiv.org/abs/1608.08624} {arXiv:1608.08624 [gr-qc]}
  \BibitemShut {NoStop}%
\bibitem [{\citenamefont {Mocz}\ \emph {et~al.}(2023)\citenamefont {Mocz},
  \citenamefont {Fialkov}, \citenamefont {Vogelsberger}, \citenamefont
  {Boylan-Kolchin}, \citenamefont {Chavanis}, \citenamefont {Amin},
  \citenamefont {Bose}, \citenamefont {Dome}, \citenamefont {Hernquist},
  \citenamefont {Lancaster}, \citenamefont {Notis}, \citenamefont {Painter},
  \citenamefont {Robles},\ and\ \citenamefont
  {Zavala}}]{https://doi.org/10.48550/arxiv.2301.10266}%
  \BibitemOpen
  \bibfield  {author} {\bibinfo {author} {\bibfnamefont {Philip}\ \bibnamefont
  {Mocz}}, \bibinfo {author} {\bibfnamefont {Anastasia}\ \bibnamefont
  {Fialkov}}, \bibinfo {author} {\bibfnamefont {Mark}\ \bibnamefont
  {Vogelsberger}}, \bibinfo {author} {\bibfnamefont {Michael}\ \bibnamefont
  {Boylan-Kolchin}}, \bibinfo {author} {\bibfnamefont {Pierre-Henri}\
  \bibnamefont {Chavanis}}, \bibinfo {author} {\bibfnamefont {Mustafa~A.}\
  \bibnamefont {Amin}}, \bibinfo {author} {\bibfnamefont {Sownak}\ \bibnamefont
  {Bose}}, \bibinfo {author} {\bibfnamefont {Tibor}\ \bibnamefont {Dome}},
  \bibinfo {author} {\bibfnamefont {Lars}\ \bibnamefont {Hernquist}}, \bibinfo
  {author} {\bibfnamefont {Lachlan}\ \bibnamefont {Lancaster}}, \bibinfo
  {author} {\bibfnamefont {Matthew}\ \bibnamefont {Notis}}, \bibinfo {author}
  {\bibfnamefont {Connor}\ \bibnamefont {Painter}}, \bibinfo {author}
  {\bibfnamefont {Victor~H.}\ \bibnamefont {Robles}}, \ and\ \bibinfo {author}
  {\bibfnamefont {Jesus}\ \bibnamefont {Zavala}},\ }\href {\doibase
  10.48550/ARXIV.2301.10266} {\enquote {\bibinfo {title} {Cosmological
  structure formation and soliton phase transition in fuzzy dark matter with
  axion self-interactions},}\ } (\bibinfo {year} {2023})\BibitemShut {NoStop}%
\bibitem [{\citenamefont {Alvarez-Rios}\ \emph {et~al.}(2025)\citenamefont
  {Alvarez-Rios}, \citenamefont {Guzm\'an},\ and\ \citenamefont
  {Niemeyer}}]{FermionBosoStars2024}%
  \BibitemOpen
  \bibfield  {author} {\bibinfo {author} {\bibfnamefont {Iv\'an}\ \bibnamefont
  {Alvarez-Rios}}, \bibinfo {author} {\bibfnamefont {Francisco~S.}\
  \bibnamefont {Guzm\'an}}, \ and\ \bibinfo {author} {\bibfnamefont {Jens}\
  \bibnamefont {Niemeyer}},\ }\bibfield  {title} {\enquote {\bibinfo {title}
  {Fermion-boson stars as attractors in fuzzy dark matter and ideal gas
  dynamics},}\ }\href {\doibase 10.1103/4tkh-7hjs} {\bibfield  {journal}
  {\bibinfo  {journal} {Phys. Rev. Lett.}\ }\textbf {\bibinfo {volume} {135}},\
  \bibinfo {pages} {161003} (\bibinfo {year} {2025})}\BibitemShut {NoStop}%
\bibitem [{\citenamefont {Alvarez-Rios}\ and\ \citenamefont
  {Guzm\'an}(2023)}]{Alvarez_Rios_2023}%
  \BibitemOpen
  \bibfield  {author} {\bibinfo {author} {\bibfnamefont {Iv\'an}\ \bibnamefont
  {Alvarez-Rios}}\ and\ \bibinfo {author} {\bibfnamefont {Francisco~S.}\
  \bibnamefont {Guzm\'an}},\ }\bibfield  {title} {\enquote {\bibinfo {title}
  {Stationary solutions of the schr\"odinger-poisson-euler system and their
  stability},}\ }\href {\doibase 10.1016/j.physletb.2023.137984} {\bibfield
  {journal} {\bibinfo  {journal} {Physics Letters B}\ }\textbf {\bibinfo
  {volume} {843}},\ \bibinfo {pages} {137984} (\bibinfo {year}
  {2023})}\BibitemShut {NoStop}%
\bibitem [{\citenamefont {Henriques}\ \emph {et~al.}(1989)\citenamefont
  {Henriques}, \citenamefont {Liddle},\ and\ \citenamefont
  {Moorhouse}}]{HENRIQUES198999}%
  \BibitemOpen
  \bibfield  {author} {\bibinfo {author} {\bibfnamefont {A.B.}\ \bibnamefont
  {Henriques}}, \bibinfo {author} {\bibfnamefont {Andrew~R.}\ \bibnamefont
  {Liddle}}, \ and\ \bibinfo {author} {\bibfnamefont {R.G.}\ \bibnamefont
  {Moorhouse}},\ }\bibfield  {title} {\enquote {\bibinfo {title} {Combined
  boson-fermion stars},}\ }\href {\doibase
  https://doi.org/10.1016/0370-2693(89)90623-0} {\bibfield  {journal} {\bibinfo
   {journal} {Physics Letters B}\ }\textbf {\bibinfo {volume} {233}},\ \bibinfo
  {pages} {99--106} (\bibinfo {year} {1989})}\BibitemShut {NoStop}%
\bibitem [{\citenamefont {Henriques}\ \emph {et~al.}(1990)\citenamefont
  {Henriques}, \citenamefont {Liddle},\ and\ \citenamefont
  {Moorhouse}}]{HENRIQUES1990511}%
  \BibitemOpen
  \bibfield  {author} {\bibinfo {author} {\bibfnamefont {A.B.}\ \bibnamefont
  {Henriques}}, \bibinfo {author} {\bibfnamefont {Andrew~R.}\ \bibnamefont
  {Liddle}}, \ and\ \bibinfo {author} {\bibfnamefont {R.G.}\ \bibnamefont
  {Moorhouse}},\ }\bibfield  {title} {\enquote {\bibinfo {title} {Stability of
  boson-fermion stars},}\ }\href {\doibase
  https://doi.org/10.1016/0370-2693(90)90789-9} {\bibfield  {journal} {\bibinfo
   {journal} {Physics Letters B}\ }\textbf {\bibinfo {volume} {251}},\ \bibinfo
  {pages} {511--516} (\bibinfo {year} {1990})}\BibitemShut {NoStop}%
\bibitem [{\citenamefont {{Ruffini}}\ and\ \citenamefont
  {{Bonazzola}}(1969)}]{Ruffini:1969}%
  \BibitemOpen
  \bibfield  {author} {\bibinfo {author} {\bibfnamefont {R.}~\bibnamefont
  {{Ruffini}}}\ and\ \bibinfo {author} {\bibfnamefont {S.}~\bibnamefont
  {{Bonazzola}}},\ }\bibfield  {title} {\enquote {\bibinfo {title} {Systems of
  self-gravitating particles in general relativity and the concept of an
  equation of state},}\ }\href {\doibase 10.1103/PhysRev.187.1767} {\bibfield
  {journal} {\bibinfo  {journal} {Phys. Rev.}\ }\textbf {\bibinfo {volume}
  {187}},\ \bibinfo {pages} {1767--1783} (\bibinfo {year} {1969})}\BibitemShut
  {NoStop}%
\bibitem [{\citenamefont {Tena-Contreras}\ \emph {et~al.}(2024)\citenamefont
  {Tena-Contreras}, \citenamefont {Alvarez-R\'ios},\ and\ \citenamefont
  {Guzm\'an}}]{CarlosIvanFranciscoGA}%
  \BibitemOpen
  \bibfield  {author} {\bibinfo {author} {\bibfnamefont {Carlos}\ \bibnamefont
  {Tena-Contreras}}, \bibinfo {author} {\bibfnamefont {Iv\'an}\ \bibnamefont
  {Alvarez-R\'ios}}, \ and\ \bibinfo {author} {\bibfnamefont {Francisco~S.}\
  \bibnamefont {Guzm\'an}},\ }\bibfield  {title} {\enquote {\bibinfo {title}
  {Construction of ground-state solutions of the gross pitaevskii poisson
  system using genetic algorithms},}\ }\href {\doibase
  10.3390/universe10080309} {\bibfield  {journal} {\bibinfo  {journal}
  {Universe}\ }\textbf {\bibinfo {volume} {10}} (\bibinfo {year} {2024}),\
  10.3390/universe10080309}\BibitemShut {NoStop}%
\bibitem [{\citenamefont {Chen}\ \emph {et~al.}(2021)\citenamefont {Chen},
  \citenamefont {Du}, \citenamefont {Lentz}, \citenamefont {Marsh},\ and\
  \citenamefont {Niemeyer}}]{ChengNiemeyer2021}%
  \BibitemOpen
  \bibfield  {author} {\bibinfo {author} {\bibfnamefont {Jiajun}\ \bibnamefont
  {Chen}}, \bibinfo {author} {\bibfnamefont {Xiaolong}\ \bibnamefont {Du}},
  \bibinfo {author} {\bibfnamefont {Erik~W.}\ \bibnamefont {Lentz}}, \bibinfo
  {author} {\bibfnamefont {David J.~E.}\ \bibnamefont {Marsh}}, \ and\ \bibinfo
  {author} {\bibfnamefont {Jens~C.}\ \bibnamefont {Niemeyer}},\ }\bibfield
  {title} {\enquote {\bibinfo {title} {New insights into the formation and
  growth of boson stars in dark matter halos},}\ }\href {\doibase
  10.1103/PhysRevD.104.083022} {\bibfield  {journal} {\bibinfo  {journal}
  {Phys. Rev. D}\ }\textbf {\bibinfo {volume} {104}},\ \bibinfo {pages}
  {083022} (\bibinfo {year} {2021})}\BibitemShut {NoStop}%
\bibitem [{\citenamefont {{Alvarez-R{\'\i}os}}\ and\ \citenamefont
  {{Guzm{\'a}n}}(2022)}]{AlvarezGuzmanMadelung}%
  \BibitemOpen
  \bibfield  {author} {\bibinfo {author} {\bibfnamefont {Iv{\'a}n}\
  \bibnamefont {{Alvarez-R{\'\i}os}}}\ and\ \bibinfo {author} {\bibfnamefont
  {Francisco~S.}\ \bibnamefont {{Guzm{\'a}n}}},\ }\bibfield  {title} {\enquote
  {\bibinfo {title} {{Construction and Evolution of Equilibrium Configurations
  of the Schr{\"o}dinger{\textendash}Poisson System in the Madelung Frame}},}\
  }\href {\doibase 10.3390/universe8080432} {\bibfield  {journal} {\bibinfo
  {journal} {Universe}\ }\textbf {\bibinfo {volume} {8}},\ \bibinfo {pages}
  {432} (\bibinfo {year} {2022})},\ \Eprint {http://arxiv.org/abs/2210.15608}
  {arXiv:2210.15608 [gr-qc]} \BibitemShut {NoStop}%
\bibitem [{\citenamefont {Guzm\'an}\ \emph {et~al.}(2021)\citenamefont
  {Guzm\'an}, \citenamefont {Alvarez-R\'ios},\ and\ \citenamefont
  {Gonz\'alez}}]{GuzmanAlvarezGonzalez2021}%
  \BibitemOpen
  \bibfield  {author} {\bibinfo {author} {\bibfnamefont {F.~S.}\ \bibnamefont
  {Guzm\'an}}, \bibinfo {author} {\bibfnamefont {I.}~\bibnamefont
  {Alvarez-R\'ios}}, \ and\ \bibinfo {author} {\bibfnamefont {J.~A.}\
  \bibnamefont {Gonz\'alez}},\ }\bibfield  {title} {\enquote {\bibinfo {title}
  {Merger of galactic cores made of ultralight bosonic dark matter},}\ }\href
  {\doibase 10.31349/revmexfis.67.75} {\bibfield  {journal} {\bibinfo
  {journal} {Rev. Mex. Fis.}\ }\textbf {\bibinfo {volume} {67}},\ \bibinfo
  {pages} {75--83} (\bibinfo {year} {2021})}\BibitemShut {NoStop}%
\bibitem [{\citenamefont {{Blum}}\ \emph {et~al.}(2025)\citenamefont {{Blum}},
  \citenamefont {{Gorghetto}}, \citenamefont {{Hardy}},\ and\ \citenamefont
  {{Teodori}}}]{Blum2025JCAP}%
  \BibitemOpen
  \bibfield  {author} {\bibinfo {author} {\bibfnamefont {Kfir}\ \bibnamefont
  {{Blum}}}, \bibinfo {author} {\bibfnamefont {Marco}\ \bibnamefont
  {{Gorghetto}}}, \bibinfo {author} {\bibfnamefont {Edward}\ \bibnamefont
  {{Hardy}}}, \ and\ \bibinfo {author} {\bibfnamefont {Luca}\ \bibnamefont
  {{Teodori}}},\ }\bibfield  {title} {\enquote {\bibinfo {title} {{Bracketing
  the soliton-halo relation of ultralight dark matter}},}\ }\href {\doibase
  10.1088/1475-7516/2025/06/050} {\bibfield  {journal} {\bibinfo  {journal}
  {JCAP}\ }\textbf {\bibinfo {volume} {2025}},\ \bibinfo {eid} {050} (\bibinfo
  {year} {2025})},\ \Eprint {http://arxiv.org/abs/2504.16202} {arXiv:2504.16202
  [astro-ph.CO]} \BibitemShut {NoStop}%
\bibitem [{\citenamefont {{\'{A}}lvarez-Rios}\ and\ \citenamefont
  {Guzm{\'{a}}n}(2022)}]{AlvarezGuzman2022}%
  \BibitemOpen
  \bibfield  {author} {\bibinfo {author} {\bibfnamefont {Iv{\'{a}}n}\
  \bibnamefont {{\'{A}}lvarez-Rios}}\ and\ \bibinfo {author} {\bibfnamefont
  {Francisco~S}\ \bibnamefont {Guzm{\'{a}}n}},\ }\bibfield  {title} {\enquote
  {\bibinfo {title} {Exploration of simple scenarios involving fuzzy dark
  matter cores and gas at local scales},}\ }\href {\doibase
  10.1093/mnras/stac3395} {\bibfield  {journal} {\bibinfo  {journal} {Monthly
  Notices of the Royal Astronomical Society}\ }\textbf {\bibinfo {volume}
  {518}},\ \bibinfo {pages} {3838--3849} (\bibinfo {year} {2022})}\BibitemShut
  {NoStop}%
\bibitem [{our()}]{ourdata}%
  \BibitemOpen
  \href {https://zenodo.org/records/18890976} {\enquote {\bibinfo {title}
  {{https://zenodo.org/records/18890976}},}\ }\BibitemShut {NoStop}%
\end{thebibliography}%

\appendix

\end{document}